\documentclass[useAMS,usenatbib]{mn2e}
\usepackage{graphicx,footnote,multirow,booktabs}
\usepackage{amssymb}
\usepackage[british]{babel}

\title[Ground-based transit observations]{Ground-based transit observations of the HAT-P-18, HAT-P-19, HAT-P-27/WASP40 and WASP-21 systems}

\author[M. Seeliger et al.]
{M.~Seeliger,$^{1}$\thanks{E-mail:martin.seeliger@uni-jena.de}
M.~Kitze,$^{1}$
R.~Errmann,$^{1,2}$
S.~Richter,$^{1}$
J.~M.~Ohlert,$^{3,4}$
W.~P.~Chen,$^{5}$
\newauthor
J.~K.~Guo,$^{5}$
E.~G\"{o}\u{g}\"{u}\c{s},$^{6}$
T.~G\"{u}ver,$^{7}$
B.~Ayd{\i}n,$^{6}$
S.~Mottola,$^{8}$
S.~Hellmich,$^{8}$
\newauthor
M.~Fernandez,$^{9}$
F.~J.~Aceituno,$^{9}$
D.~Dimitrov,$^{10}$
D.~Kjurkchieva,$^{11}$
E.~Jensen,$^{12}$
\newauthor
D.~Cohen,$^{12}$
E.~Kundra,$^{13}$
T.~Pribulla,$^{13}$
M.~Va\v{n}ko,$^{13}$
J.~Budaj,$^{14,13}$
M.~Mallonn,$^{15}$
\newauthor
Z.-Y.~Wu,$^{16}$
X.~Zhou,$^{16}$
St.~Raetz,$^{1,17}$
C.~Adam,$^{1}$
T.~O.~B.~Schmidt,$^{1}$
A.~Ide,$^{1}$
\newauthor
M.~Mugrauer,$^{1}$
L.~Marschall,$^{19}$
M.~Hackstein,$^{20}$
R.~Chini,$^{20,21}$
M.~Haas,$^{20}$
T.~Ak,$^{7}$
\newauthor
E.~G\"uzel,$^{22}$
A.~\"Ozd\"onmez,$^{23}$
C.~Ginski,$^{1,24}$
C.~Marka,$^{1}$
J.~G.~Schmidt,$^{1}$
B.~Dincel,$^{1}$
\newauthor
K.~Werner,$^{1}$
A.~Dathe,$^{1}$
J.~Greif,$^{1}$
V.~Wolf,$^{1}$
S.~Buder,$^{1}$
A.~Pannicke,$^{1}$
\newauthor
D.~Puchalski$^{25}$
and R.~Neuh\"auser$^{1}$\\
$~^{1}$Astrophysical Institute and University Observatory Jena, Schillerg\"a\ss{}chen 2-3, D-07745 Jena, Germany\\
$~^{2}$Abbe Center of Photonics, Friedrich Schiller Universit\"at, Max-Wien-Platz 1, D-07743 Jena, Germany\\
$~^{3}$Astronomie Stiftung Trebur, Michael Adrian Observatorium, Fichtenstra\ss{}e 7, D-65468 Trebur, Germany\\
$~^{4}$University of Applied Sciences, Technische Hochschule Mittelhessen, D-61169 Friedberg, Germany\\
$~^{5}$Graduate Institute of Astronomy, National Central University, Jhongli City, Taoyuan County 32001, Taiwan (R.O.C.)\\
$~^{6}$Sabanci University, Orhanli-Tuzla 34956, Istanbul, Turkey\\
$~^{7}$Faculty of Sciences, Department of Astronomy and Space Sciences, Istanbul University, 34119 \.Istanbul, Turkey\\
$~^{8}$Deutsches Zentrum f\"ur Luft- und Raumfahrt e.V., Institut f\"ur Planetenforschung, Rutherfordstr. 2, D-12489 Berlin, Germany\\
$~^{9}$Instituto de Astrofisica de Andalucia, CSIC, Apdo. 3004, E-18080 Granada, Spain\\
$^{10}$Institute of Astronomy and NAO, Bulgarian Academy of Sciences, 72 Tsarigradsko Chaussee Blvd., 1784 Sofia, Bulgaria\\
$^{11}$Shumen University, 115 Universitetska str., 9700 Shumen, Bulgaria\\
$^{12}$Department of Physics and Astronomy, Swarthmore College, Swarthmore, PA 19081, USA\\
$^{13}$Astronomical Institute, Slovak Academy of Sciences, 059 60, Tatransk\'{a} Lomnica, Slovakia\\
$^{14}$Research School of Astronomy and Astrophysics, Australian National University, Canberra, ACT 2611, Australia\\
$^{15}$Leibnitz-Institut f\"ur Astrophysik Potsdam, An der Sternwarte 16, D-14482 Potsdam, Germany\\
$^{16}$Key Laboratory of Optical Astronomy, NAO, Chinese Academy of Sciences, 20A Datun Road, Beijing 100012, China\\
$^{17}$European Space Agency, ESTEC, SRE-S, Keplerlaan 1, 2201AZ Noordwijk, The Netherlands\\
$^{19}$Gettysburg College Observatory, Department of Physics, 300 North Washington St., Gettysburg, PA 17325, USA\\
$^{20}$Astronomisches Institut, Ruhr-Universit\"at Bochum, Universit\"atsstra\ss{}e 150, D-44780 Bochum, Germany\\
$^{21}$Instituto de Astronom{\'i}a, Universidad Cat\'{o}lica del Norte, Avenida Angamos 0610, Casilla 1280 Antofagasta, Chile\\
$^{22}$Faculty of Science, Department of Astronomy and Space Sciences, Univserity of Ege, Bornova, 35100 \.Izmir, Turkey\\
$^{23}$Graduate School of Science and Engineering, Department of Astronomy and Space Sciences, Istanbul University, 34116 Istanbul, Turkey\\
$^{24}$Sterrewacht Leiden, P.O. Box 9513, Niels Bohrweg 2, 2300RA Leiden, The Netherlands\\
$^{25}$Centre for Astronomy, Faculty of Physics, Astronomy and Informatics, N. Copernicus University, Grudziadzka 5, PL-87-100 Toru\'{n}, Poland}

\date{}

\pagerange{\pageref{firstpage}--\pageref{lastpage}} \pubyear{2013}

\begin{document}

\label{firstpage}

\maketitle

\begin{abstract}
As part of our ongoing effort to investigate transit timing variations (TTVs) of known exoplanets, we 
monitored transits of the four exoplanets HAT-P-18b, HAT-P-19b, HAT-P-27b/WASP-40b and WASP-21b.
All of them are suspected to show TTVs due to the known properties of their host systems based on the respective 
discovery papers. 
During the past three years 42 transit observations were carried out, mostly using telescopes of the Young Exoplanet
Transit Initiative. 
The analyses are used to refine the systems orbital parameters. 
In all cases we found no hints for significant TTVs, or changes in the system 
parameters inclination, fractional stellar radius and planet to star radius ratio and thus could confirm the already published results.
\end{abstract}


\section{Introduction}
\label{sec:Introduction}

Observing extra solar planets transiting their host stars has become an important tool for planet 
detection and is used to obtain and constrain fundamental system parameters: The inclination has to be close to $90^\circ$, while
the planet to star radius ratio is constrained mainly by the transit depth. Also, in combination with spectroscopy, the semimajor axis
and the absolute planet and star radii can be obtained as well.

Several years ago, when the first results of the \textit{Kepler} mission were published (see 
\citealt{Keplerdata} for first scientific results, \citealt{Keplerinstrument} for an instrument description),
studying the transit timing became one of the standard techniques in the analysis of transit observations.
Since space based missions are able to observe many consecutive transit
events with a high precision, one can detect even small variations of the transit intervals
indicating deviations from a strictly Keplerian motion and thus yet hidden planets in the observed system.
Furthermore, with the discovery of multi-planetary systems, transit timing
variations (TTVs) are used to find the masses of the companions without the need of 
radial velocity measurements due to the influence of planetary interaction on TTVs.
Since many planet candidates found in photometric surveys are too faint
for radial velocity follow-up even with bigger telescopes, TTV analyses can be considered as a photometric 
work-around to estimate masses.

Although the existence of TTVs can be shown in already known exoplanetary systems, only a few 
additional planet candidates have been found using TTVs so far (for recent examples of a proposed
additional body indicated by TTV analyses see e.g. \citealt{Maciejewski2011a} and \citealt{vanEylen2014}). This is not suprising, since large 
bodies often can be found using radial velocity measurements or direct transit detections, while small
(e.g. Earth-like) objects result in small TTV amplitudes and therefore high precision timing measurements are needed. 
These measurements can already be acquired with medium size ground-based telescopes.

Commonly the transit mid-time of each observation is plotted into an observed minus
calculated ({O--C}) diagram \citep{FordHolman2007}, where the difference between the observed transit mid-time
and the mid-time obtained using the initial ephemeris is shown versus the observing epoch. In such a diagram remaining slopes indicate a wrong orbital period, while
e.g. periodic deviations from a linear trend indicate perturbing forces.

Besides the discovery of small planets, the amount of known massive planets on close-in
orbits increased as well. 
First studies on a larger sample of planet candidates detected with \textit{Kepler} suggest that hot giant planets exist in single planet systems only \citep{Steffen2012}. 
However, \citet{Szabo2013} analysed a larger sample of \textit{Kepler} hot Jupiters and found a few cases where TTVs can not be explained by other causes (e.g.
artificial sampling effects due to the observing cadence) but the existence of perturbers -- additional planets or even exo-moons -- in the respective system. 
In addition \citet{Szabo2013} point toward the 
planet candidates KOI-338, KOI-94 and KOI-1241 who are all hot Jupiters in multi-planetary systems,
as well as the HAT-P-13 system with a hot Jupiter accompanied by
a massive planet on an eccentric outer orbit \citep[see also][]{Szabo2010} and the WASP-12 system with a proposed companion candidate found by TTV analysis \citep{Maciejewski2011a}.

The origin of those planets is yet not fully understood. One possible formation scenario 
shows that close-in giant planets could have migrated inwards 
after their creation further out \citep{Steffen2012}. In that case, inner and close outer planets 
would have either been thrown out of the system, or caught in resonance. 
In the latter case, even small perturbing masses, e.g. Earth-mass objects, can result in TTV amplitudes 
in the order of several minutes (see \citealt{FordHolman2007} or \citealt{Seeliger2014}).
Though \textit{Kepler} is surveying many of those systems, it is 
necessary to look at the most promising candidates among all close-in giant planets discovered so far,
including stars outside the field of view of \textit{Kepler}.
In our ongoing study\footnotemark\footnotetext{see http://ttv.astri.umk.pl/doku.php
for a project overview} of 
TTVs in exoplanetary systems we perform photometric follow-up observations of specific promising 
transiting planets where additional bodies are expected. 
The targets are selected by the following criteria: 

\begin{itemize}
	\vspace{-0.4em}
	\setlength{\itemsep}{1pt}
	\setlength{\parskip}{0pt}
	\setlength{\parsep}{0pt}
	\item[(i)] The orbital solution of the known transiting planet shows non-zero 
	eccentricity (though the circularization time-scale is much shorter than the 
	system age) and/or deviant radial velocity (RV) data points -- both possibly indicating a perturber.
	\item[(ii)] The brightness of the host star is $V\leq13\:$mag and the transit depth 
	is at least $10\:$mmag to ensure sufficient 
	photometric and timing precision at 1-2m class ground-based telescopes.
	\item[(iii)] The target is visible from the Northern hemisphere.
	\item[(iv)] The target has not been studied for TTV signals before.
\end{itemize}

In the past the transiting exoplanets WASP-12b \citep{Maciejewski2011a, Maciejewski2013a}, 
WASP-3b \citep{Maciejewski2010,Maciejewski2013b}, 
WASP-10b \citep[][Maciejewski et al. 2014 in prep.]{Maciejewski2011b}, 
WASP-14b \citep{Raetz2012}, TrES-2 \citep{Raetz2014}, and 
HAT-P-32b \citep{Seeliger2014} have been studied by our group in detail. 
In most cases, except for WASP-12b, no TTVs could be confirmed.

Here, we extend our investigations to search for TTVs in the HAT-P-18, HAT-P-19, HAT-P-27/WASP-40 and WASP-21 planetary systems.
In Section~\ref{sec:Targets} we give a short description of the targets analysed within this project.
Section~\ref{sec:DataAquisitionAndReduction} explains the principles
of data acquisition and reduction and gives an overview of the telescopes used for observation.
The modeling procedures are described in Section~\ref{sec:Analyses}, followed by the results 
in Section~\ref{sec:Results}. Finally, Section~\ref{sec:Summary} gives a summary of our project.

\begin{table*}
	\centering
	\caption{The observing telescopes that gathered data within the TTV project 
	for HAT-P-18b, HAT-P-19b, HAT-P-27b/WASP-40b and WASP-21b in order 
	of the number of observed transit events of the Observatory. The table lists the 
	telescopes and corresponding observatories, as well as the telescope diameters 
	$\oslash$ and number of observed transit events per telescope in this project 
	N$_{\textrm{tr}}$.}\label{tab:Observatories}
	{\setlength{\tabcolsep}{4pt}
	\begin{tabular}{llllc}
	\toprule
	\# & Observatory                                           & Telescope (abbreviation)                      & $\oslash \left(m\right)$ & $N_{\textrm{tr}}$\\
	\midrule
	1  & Michael Adrian Observatory Trebur (Germany)            & T1T (Trebur 1.2m)                            & 1.2               & 8 \\ 
	2  & Graduate Institute of Astronomy Lulin (Taiwan \& USA)  & Tenagra II (Tenagra 0.8m)                    & 0.8               & 5 \\
	   &                                                        & RCOS16 (Lulin 0.4m)                          & 0.4               & 2 \\
	3  & University Observatory Jena (Germany)                  & 90/60 Schmidt (Jena 0.6m)                    & 0.9/0.6           & 4 \\
	   &                                                        & Cassegrain (Jena 0.25m)                      & 0.25              & 3 \\
	4  & T\"UB\.ITAK National Observatory (Turkey)              & T100 (Antalya 1.0m)                          & 1.0               & 5 \\
	5  & Calar Alto Astronomical Observatory (Spain)            & 1.23m Telescope (CA-DLR 1.2m)                & 1.23              & 4 \\
	6  & Sierra Nevada Observatory (Spain)                      & Ritchey-Chr\'{e}tien (OSN 1.5m)              & 1.5               & 2 \\
	7  & Peter van de Kamp Observatory Swarthmore (USA)         & RCOS (Swarthmore 0.6m)                       & 0.6               & 2 \\
	8  & National Astronomical Observatory Rozhen (Bulgaria)    & Ritchey-Chr\'{e}tien-Coud\'{e} (Rozhen 2.0m) & 2.0               & 1 \\
	   &                                                        & Cassegrain (Rozhen 0.6m)                     & 0.6               & 1 \\
	9  & Teide Observatory, Canarian Islands (Spain)            & STELLA-I (Stella 1.2m)                       & 1.2               & 2 \\
	10 & University Observatory Bochum (Cerro Armazones, Chile) & VYSOS6 (Chile 0.15m)                         & 0.15              & 1 \\
	11 & Xinglong Observing Station (China)                     & 90/60 Schmidt (Xinglong 0.6m)                & 0.9/0.6           & 1 \\
	12 & Gettysburg College Observatory (USA)                   & Cassegrain (Gettysburg 0.4m)                 & 0.4               & 1 \\
	13 & Star\'{a} Lesn\'{a} Observatory (Slovak Rep.)          & 0.5m Reflector (StaraLesna 0.5m)             & 0.5               & 1 \\
	14 & Istanbul University Telescope at \c{C}anakkale (Turkey)& 0.6m Telescope        (\c{C}anakkale 0.6m)   & 0.6               & 1 \\ 
	15 & Toru\'{n} Centre for Astronomy (Poland)                & 0.6m Cassegrain Telescope  (Tor\'{u}n 0.6m)  & 0.6               & 1 \\
	\bottomrule
	\end{tabular}}
\end{table*}

\section{Targets}\label{sec:Targets}

\subsection{HAT-P-18b and HAT-P-19b}\label{subsec:H18H19}
\citet{HatP18u19} reported on the discovery of the exoplanets HAT-P-18b and HAT-P-19b. 
The two Saturn-mass planets orbit their early K type stars with periods of 5.51d and 
4.01d, respectively. 

In case of HAT-P-18b \citet{HatP18u19} found the eccentricity to be slightly non-zero
$\left(e=0.084\pm0.048\right)$. Recent studies of \citet{Esposito2014} 
found the eccentricity to be consistent with a non-eccentric retrograde orbit by analysing the 
Rossiter-McLaughlin effect. \citet{Knutson2013} also analysed the RV signal and found a 
jitter in the order of $17.5\:$m\,s$^{-1}$ that remains unexplained.
Interestingly, the transit data listed in the exoplanet transit database 
\citep[ETD;][]{Podanny2010} shows a huge spread in the transit depth in the 
order of several tens of milli-magnitudes. Thus we included HAT-P-18b
in our list of follow-up objects to also confirm or refute these transit depth variations.
In addition, we performed a monitoring project for HAT-P-18 over a longer time span to
possibly find overall brightness variations. In \citet{Ginski2012} we used lucky imaging with
Astralux at the Calar Alto 2.2m Telescope to search for additional low-mass stellar 
companions in the system. With the data gathered in this previous study we could already exclude 
objects down to a mass of $0.140\pm0.022\,\textrm{M}_{\odot}$ at angular separations
as small as 0.5\,arcsec and objects down to 0.099$\pm$0.008\,M$_\odot$ outside of 2\,arcsec.

For HAT-P-19b a small eccentricity of $e=0.067\pm0.042$ was determined by 
\citet{HatP18u19}. They also found a linear trend in the RV residuals pointing 
towards the existence of a long period perturber in the system. 
Within this project we 
want to address the problem of the proposed perturber using photometric methods, i.e. 
follow-up transit events to find planetary induced TTV signals.

\subsection{HAT-P-27b/WASP-40b}\label{subsec:H27}

HAT-P-27b \citep{HatP27}, independently discovered as WASP-40b by \citet{Wasp40} within
the WASP-survey \citep{Pollacco2006}, is a typical hot Jupiter with a period of 3.04d. 
While the eccentricity was found to be $e=0.078\pm0.047$ by \citet{HatP27}, 
\citet{Wasp40} adopted a non-eccentric orbit. However, the latter author found a huge spread 
in the RV data with up to $40\:$m\,s$^{-1}$ deviation from the circular single planet 
solution. According to \citet{Wasp40} one possible explanation, despite a changing
activity of the K-type host star, is the existence of a perturber that might 
not be seen in the \citet{HatP27} data due to the limited data set.
However, the authors suggest further monitoring to clearify the nature of the system.
One possibility is to study the companion hypothesis from the TTV point of view.

Another interesting aspect of HAT-P-27b is the transit shape which \citet{HatP27}
fitted using a flat bottom model. \citet{Wasp40} and \citet{Sada} found the transit rather to have a roundish shape. 
From the grazing criterion \citep{Smalley2011} they concluded that the system is probably grazing, which
would explain the unusual shape of the transit.
However it is not clear why this is not seen in the \citet{HatP27} data, hence
it is still not clear which shape is real.

\subsection{WASP-21b}
The planetary host star WASP-21, with its Saturn-mass planet on a 4.32d orbit discovered by \citet{Wasp21}, is one the 
most metal-poor planet hosts accompanied by one of the least dense planets
discovered by ground-based transit searches to date. 
\citet{Wasp21} found that including a small non-zero eccentricity to the fit does not improve the
results. Hence, they concluded that the eccentricity is consistent with zero. 

However, in a later study \citet{Barros2011}
found the G3V star to be in the process of moving off the main sequence. 
Thus, we included further observations of WASP-21b planetary transits to improve the knowledge on 
this system.

\section{Data acquisition and reduction}
\label{sec:DataAquisitionAndReduction}

\begin{table}
	\centering
	\caption{The list of all transit observations gathered within the TTV project sorted 
	by object and date. Though no preselections for quality or completeness have been applied to this list, 
	transits used for further analysis have been marked by an asterisk. The
	filter subscripts B, C and J denote the photometric systems of Bessel, Cousins and Johnson,
	respectively. The last column lists the number of exposures and the exposure time of each observation.
	}\label{tab:AllObservation}
	\begin{tabular}{llllr@{\,x\,}r}
	\toprule
	\# & Date & Telescope                                  & Filter & \multicolumn{2}{c}{Exposures}\\
	\midrule
	\multicolumn{6}{c}{\emph{HAT-P-18b}}\\
	\midrule
	1*  & 2011-04-21 &  Trebur 1.2m                         &  R$_B$ &  189 & 90\,s       \\
	2   & 2011-05-02 &  Trebur 1.2m                         &  R$_B$ &  123 & 45\,s       \\
	3*  & 2011-05-24 &  Trebur 1.2m                         &  R$_B$ &  323 & 60\,s       \\
	4   & 2011-06-04 &  Rozhen 2.0m                         &  R$_C$ &  1000& 10\,s       \\
	5*  & 2012-05-05 &  Rozhen 0.6m                         &  I$_C$ &  219 & 90\,s       \\
	6   & 2012-06-07 &  CA DLR 1.23m                        &  B$_J$ &  250 & 60\,s       \\
	7   & 2013-04-28 &  Antalya 1.0m                        &  R     &  214 &  50\,s       \\
	8   & 2014-03-30 &  Toru\'{n}  0.6m                     &  clear &  297 & 40\,s\\
	\midrule
	\multicolumn{6}{c}{\emph{HAT-P-19b}}\\
	\midrule
	9*  & 2011-11-23 &  Jena 0.6m                           & R$_B$  &  246& 50\,s       \\
	10  & 2011-11-23 &  Jena 0.25m                          & R$_B$  &  320& 50\,s      \\
	11  & 2011-11-23 &  Trebur 1.2m                         & R$_B$  &  461& 30\,s      \\
	12  & 2011-12-05 &  Jena 0.6m                           & R$_B$  &  129& 60\,s      \\
	13  & 2011-12-05 &  Jena 0.25m                          & V$_B$  &  28 &300\,s      \\
	14* & 2011-12-09 &  Jena 0.6m                           & R$_B$  &  290& 50\,s      \\
	15  & 2011-12-09 &  Jena 0.25m                          & V$_B$  &  118&150\,s      \\
	16  & 2011-12-09 &  Trebur 1.2m                         & R$_B$  &  380& 35\,s      \\
	17* & 2011-12-17 &  CA DLR 1.23m                        & R$_J$  &  273& 60\,s       \\
	18  & 2014-08-01 &  Antalya 1.0m                        & R      &  148 & 60\,s\\
	19  & 2014-08-05 &  Antalya 1.0m                        & R      &  196 & 40\,s\\
	20  & 2014-08-21 &  Jena 0.6m                           & R$_B$  &   152&50\,s\\
	21* & 2014-10-04 &  Jena 0.6m                           & R$_B$  &   280&50\,s\\
	\midrule
	\multicolumn{6}{c}{\emph{HAT-P-27b}}\\
	\midrule
	22* & 2011-04-05 &  Lulin 0.4m                          & R$_B$  &  166&40\,s         \\
	23* & 2011-04-08 &  Lulin 0.4m                          & R$_B$  &  250&40\,s         \\
	24  & 2011-05-03 &  Stella 1.2m                         & H$\alpha$ &180& 100\,s      \\
	25* & 2011-05-05 &  Trebur 1.2m                         & R$_B$  &  162&70\,s        \\
	26  & 2011-05-08 &  Stella 1.2m                         & H$\alpha$ &190&100\,s      \\
	27  & 2011-05-21 &  Tenagra 0.8m                        & R      &  141&40\,s        \\
	28  & 2012-03-07 &  StaraLesna 0.5m                     & R      &  361&30\,s        \\
	29  & 2012-03-29 &  Tenagra 0.8m                        & R      &  240&30\,s        \\
	30  & 2012-04-01 &  Tenagra 0.8m                        & R      &  329&20\,s        \\
	31  & 2012-04-04 &  Tenagra 0.8m                        & R      &  333&20\,s        \\
	32  & 2012-04-25 &  Xinglong 0.6m                       & R      &  154&40\,s        \\
	33  & 2012-05-16 &  Trebur 1.2m                         & R$_B$  &  231&70\,s        \\
	34  & 2012-05-25 &  Chile 0.15m                         &I$_J$/R$_J$& 220&80\,s     \\
	35  & 2012-06-13 &  Tenagra 0.8m                        & R      &  223&15\,s        \\
	36* & 2013-06-03 &  Antalya   1.0m                      & R      &  156&60\,s        \\
	37* & 2013-06-03 &  OSN 1.5m                            & R      &  435&30\,s        \\
	38  & 2013-06-06 &  CA DLR 1.23m                        & R$_J$  &  172&60\,s        \\
	39* & 2014-06-18 &  Antalya 1.0m                        & R      &  146&50\,s        \\
	\midrule
	\multicolumn{6}{c}{\emph{WASP-21b}}\\
	\midrule
	40* & 2011-08-24 &  Swarthmore 0.6m                     & R$_B$  &  545&45\,s        \\
	41  & 2011-08-24 &  Gettysburg 0.4m                     & R      &  230&60\,s        \\
	42* & 2012-08-16 &  Trebur 1.2m                         & R$_B$  &  365&40\,s        \\
	43  & 2012-10-20 &  Antalya   1.0m                      & R      &  242&40\,s        \\
	44* & 2013-09-18 &  CA DLR 1.23m                        & R$_J$  &  584&30\,s        \\
	45  & 2013-09-22 &  Antalya 1.0m                        & R      &  208&50\,s        \\
	46  & 2013-09-22 &  Ulupinar 0.6m                       & R$_B$  &  163&110\,s       \\
	\bottomrule
	\end{tabular}
\end{table}

Our observations make use of YETI network telescopes
\citep[Young Exoplanet Transit Initiative;][]{YETI}, a worldwide network of small to 
medium sized telescopes mostly on the Northern hemisphere established to explore 
transiting planets in young open clusters.

A summary of all participating telescopes and the number of performed observations can be found in Table~\ref{tab:Observatories}.
Most of the observing telescopes are part of the YETI network. This includes telescopes at Cerro Armazones (Chile,
operated by the University of Bochum), Gettysburg (USA), Jena (Germany), Lulin (Taiwan), Rozhen 
(Bulgaria), Sierra Nevada (Spain), Star\'{a} Lesn\'{a} (Slovak Republic), Swarthmore (USA), Tenagra (USA,
operated by the National Central University of Taiwan) and Xinglong (China). For details about 
location, mirror and chip see \citet{YETI}.

In addition to the contribution of the YETI telescopes, we obtained data using the following telescopes:
\begin{itemize}
	\item the 1.2m telescope of the German-Spanish Astronomical Center on Calar Alto (Spain), which 
	is operated by German Aerospace Center (DLR).
	\item the 1.2m robotic telescope STELLA-I, situated at Teide Observatory on Tenerife (Spain) and operated
	by the Leibnitz-Institut f\"ur Astrophysik Potsdam (AIP).
	\item the Trebur 1Meter Telescope operated at the Michael Adrian Observatory Trebur (Germany)
	\item the T100 telescope of the T\"UBITAK National Observatory (Turkey)
	\item the 0.6m telescope (CIST60) at Ulup{\i}nar Observatory operated by Istanbul University (Turkey)
	\item the 0.6m Cassegrain telescope of the Toru\'{n} Centre for Astronomy (Poland)
\end{itemize}

Besides the transit observations, the Jena 0.6m telescope with its Schmidt Teleskop Kamera \citep{STK} was used to perform a long 
term monitoring of HAT-P-18 as described in Sections~\ref{subsec:H18H19} and \ref{sec:ResultsH18}.

Between 2011 April and 2013 June our group observed 45 transit events (see 
Table~\ref{tab:AllObservation}) using 18 different telescopes (see 
Table~\ref{tab:Observatories}). 15 observations could be used for 
further analysis, while 30 observations had to be rejected due to several reasons, e.g. 
no full transit event has been observed or bad weather conditions and hence low signal to noise.
E.g. \cite{Southworth2009a,Southworth2009b} showed that defocusing the telescope allows to 
reduce atmospheric and flat fielding effects. 
Since a defocused image spreads the light over several CCD pixel, one can increase the exposure time and hence
the effective duty cycle of the CCD assuming a constant read out time
\citep[as mentioned also in the conclusions of][]{Barros2011}. Thus we tried to defocuse 
the telescope and increase the exposure time during all our observations. 
Table~\ref{tab:AllInputValues} lists the ingress/egress durations $\tau$ 
derived using the formula (18) and (19) given in \citet{Winn2010}. With our strategy we obtain at least one 
data point within 90s. This ensures to have at least 10 data points during ingress/egress 
phase which is required to fit the transit model to the data and get precise transit 
mid-times. 

\begin{table*}
	\centering
	\caption{The input parameters for the JKTEBOP \& TAP runs for all objects listed 
	in Section~\ref{sec:Targets}. All values have been obtained from the original discovery papers.
	LD coefficients are taken from 
	\citet{Claret2011} linear interpolated in terms of $T_{\textrm{eff}}$, $\log g$ and 
	$\left[ \textrm{Fe}/\textrm{H}\right]$ using the \textit{EXOFAST/QUADLD} code  
	\citep{Eastman2013}. Free parameters are marked by an asterisk.
	At the bottom the duration of ingress and egress according to \citet{Winn2010} has been added.
	}\label{tab:AllInputValues}
	\begin{tabular}{lr@{$.$}lr@{$.$}lr@{$.$}lr@{$.$}lr@{$.$}lr@{$.$}l}
	\toprule
	Object 									& \multicolumn{2}{c}{HAT-P-15b}	 & \multicolumn{2}{c}{HAT-P-18b} & \multicolumn{2}{c}{HAT-P-19b} & \multicolumn{2}{c}{HAT-P-27b} & \multicolumn{2}{c}{WASP-21b} & \multicolumn{2}{c}{WASP-38b} \\
	\midrule
	$r_\textrm{p}+r_\textrm{s}$*			&	 0&0575(19)					&  0&0575(19)					&  0&0709(33)					& 0&1159(65)					& 0&0959(44)					& 0&0829(27)\\
	$R_\textrm{p}/R_\textrm{s}$*			&	 0&1019(09) 				&  0&1365(15)					&  0&1418(20)					& 0&1186(31)					& 0&1040(35)					& 0&0844(21)\\
	$i$ $(^\circ)$*							&	89&1(2) 					& 88&8(3) 						& 88&2(4)						& 84&7(7)						& 88&75(84)						& 89&69(30)\\
	$a/R_\textrm{s}$*						&	19&16(62)					& 16&04(75)						& 12&24(67)						& 9&65(54)						& 10&54(48)						& 12&15(39)\\
	$M_\textrm{p}/M_\textrm{s}$				&	 0&0018(01) 				&  0&000243(26)					&  0&000329(37)					& 0&000663(58)					& 0&000282(19)					& 0&00213(11)\\
	$e$										&	 0&190(19) 					&  0&084(48)					&  0&067(42)					& 0&078(47)						& \multicolumn{2}{c}{0}			& 0&0314(46)\\
	$P$ (d)									&	10&863502(27)				&  5&5080023(06)				&  4&008778(06)					& 3&039586(12)					& 4&322482(24)					& 6&871815(45)\\
	\midrule
	$R$ (mag)								&	11&81						& 12&61							& 12&82							& 11&98							& 11&52							& 9&22\\
	$T_{\textrm{eff}}$ (K)					&	\multicolumn{2}{c}{5568(90)}& \multicolumn{2}{c}{4803(80)}	& \multicolumn{2}{c}{4990(130)}	& \multicolumn{2}{c}{5300(90)}	&\multicolumn{2}{c}{5800(100)}	&\multicolumn{2}{c}{6150(80)}\\
	$\log g$ (cgs)							&	 4&38(03)					&  4&57(04)						&   4&54(05)					&  4&51(04)						&  4&2(1)						&  4&3(1)\\
	$[\textrm{Fe}/\textrm{H}]$ (dex)		&	+0&22(08)					& +0&10(08)						&  +0&23(08)					& +0&29(10)						& -0&46(11)						& -0&12(07)\\
%	$[\textrm{M}/\textrm{H}]$ (dex)			&	+0&18(11) 					& +0&08(11)						&  +0&19(11)					& +0&24(14)						& -0&37(16)						& -0.10&(10)\\
	$v \sin i$ (km s$^{-1}$)				&	 2&0(5) 					&  0&5(5)						&   0&7(5)						&  0&4(4)						&  1&5(6)						&   8&6(4)\\
	\midrule
	LD law of the star						& \multicolumn{2}{l}{quadratic}	& \multicolumn{2}{l}{quadratic}	& \multicolumn{2}{l}{quadratic}	& \multicolumn{2}{l}{quadratic}	& \multicolumn{2}{l}{quadratic} & \multicolumn{2}{l}{quadratic}\\
	$R$ band linear*						&	0&4200						& 0&5736						& 0&5433						& 0&4808						& 0&3228						& 0&2998 			\\
	$R$ band non-linear*					&	0&2525						& 0&1474						& 0&1710						& 0&2128						& 0&2982						& 0&3095 			\\
	$V$ band linear*						&	0&5274						& 0&7180						& 0&6783						& 0&6002						& 0&4055						& 0&3834 			\\
	$V$ band non-linear*					&	0&2164						& 0&0697						& 0&1039						& 0&1643						& 0&2892						& 0&2992 			\\
	\midrule
	$\tau_{\mathrm{egress/ingress}}$ (min)	&	27&8						& 22&8							& 23&1							& 37&8							& 20&1							& 21&9 			\\
	\bottomrule
	\end{tabular}
\end{table*}

All data has been reduced in a standard way by applying dark/bias and flat field 
corrections using \textsc{iraf}\footnotemark\footnotetext{\textsc{iraf} is distrubuted
by the National Optical Astronomy Observatories, which are operated by the Association 
of Universities for Research in Astronomy, Inc., under coorporative agreement with 
the National Science Foundation.}. The respective calibration images have been obtained 
in the same night and with the same focus as the scientific observations. This is necessary
especially if the pointing of the telescope is not stable. When using calibration images 
obtained with different foci, patterns remain in the images that lead to distortions in
the light curve.

Besides our own observations, we also use literature data. This involves data
from the respective discovery papers mentioned in Section~\ref{sec:Targets}, as well as data from 
\citet{Esposito2014} in case of HAT-P-18b, \citet{Sada} in case of HAT-P-27b, 
\citet{Barros2011} and \citet{Ciceri2013} for WASP-21b, and \citet{Simpson2010} for WASP-38b.

\section{Analyses}
\label{sec:Analyses}

The light curve extraction and modelling is performed analogous to the procedure 
described in detail in \citet{Seeliger2014}.

\subsection{Light curve extraction}

The Julian date of each image is calculated from the header information of the start of the exposure and the exposure time.
To precisely determine the mid-time of the transit event these informations have to be stored most accurate.
The reliability of the final light curve model thus also depends on a precise time
synchronization of the telescope computer system. Observing transits with multiple telescopes at the same time enables to look for synchronization errors which would
otherwise lead to artifacts in the {O--C} diagram (as shown for HAT-P-27b in epoch 415, see Section~\ref{subsec:ergH27}).

We use differential aperture photometry to extract the light curve from the reduced images
by measuring the brightness of all bright stars in the field with routines provided by 
\textsc{iraf}. The typical aperture radius is $\approx1.5$ times the mean full width half 
maximum of all stars in the field of view (FoV). The best fitting aperture is found by manually 
varying the aperture radius by a few pixels to minimize the photometric scatter. 
The final light curve is created by comparing the star of interest against an artificial 
standard star composed of the (typically 15-30) brightest stars in the FoV weighted by their respective constantness
as introduced by \citet*{Broeg2005}.

The final photometric errors are based on the instrumental \textsc{iraf} measurement errors. The error of the constant
comparison stars are rescaled by their photometric scatter using shared scaling factors in order to achieve a mean $\chi_{red}^2\approx1$ for all comparison stars. The error bars
of the transit star are rescaled afterwards using the same scaling factors \citep[for further details on the procedure see][]{Broeg2005}.

Due to atmospheric and air-mass effects transit light curves show trends. They can impact
the determination of transit parameters. To eliminate such effects we start the observation about 1 hour
before and finish about 1 hour after the transit itself. Thus we can detrend the observations
by fitting a second order polynomial to the out-of-transit data.

\subsection{Modelling with TAP and JKTEBOP}

To model the light curves we used the Transit Analysis Package \citep[\textsc{tap};][]{TAP}.
The modelling of the transit light curve is done by using the 
\textsc{exofast} routines \citep*{Eastman2013} with the light curve model of 
\citet{MandelAgol2002}. For error estimation \textsc{tap} uses Markov Chain Monte Carlo simulations (in our case 10 times 10$^5$ MCMC chains)
together with wavelet-based likelihood functions \citep{CarterWinn2009}. 
The coefficients for the quadratic limb darkening (LD) law used by \textsc{tap} are taken 
from the \textsc{exofast/quadld}-routine of \citet{Eastman2013}\footnotemark 
\footnotetext{the limb darkening calculator is available online at 
http://astroutils.astronomy.ohio-state.edu/exofast/limbdark.shtml}
that linearly interpolates the LD tables of \citet{Claret2011}.

For comparison we also use \textsc{jktebop} \citep[see][and references therein]{JKTEBOP} 
which is based on the \textsc{ebop} code for eclipsing binaries \citep{Etzel1981,Popper1981}.
To compare the results with those obtained with \textsc{tap}, we only use a quadratic LD
law which is sufficient for ground-based data. 
For error estimation 
we used Monte Carlo simulations ($10^4$ runs), bootstrapping ($10^4$ data sets), and a 
residual shift method as provided by \textsc{jktebop}.

As input values we use the system properties presented in the respective discovery papers
(see Table~\ref{tab:AllInputValues} for a summary).
For both light curve fitting procedures we took the original light curve as well as a
threefold binned one. Though the binned light curves result in a lower rms of the fit, 
no better timing result can be achieved due to a longer cadence.

As free parameters we use the mid-transit time $T_\textrm{mid}$, inclination $i$, and planet to star radius ratio 
$k=r_\textrm{p}/r_\textrm{s}$ (with $r_\textrm{p}$ and $r_\textrm{s}$ being the planet 
and stellar radius scaled by the semimajor axis, respectively). 
In case of \textsc{tap} the inverse fractional stellar radius $a/R_\textrm{s}=1/r_\textrm{s}$, in case of 
\textsc{jktebop} the sum of the fractional radii $\left(r_\textrm{p}+r_\textrm{s}\right)$ 
is fitted as well. Both quantities are an expression of the transit duration and can 
be transformed into each other according to the following equation:
\[
	a/R_\textrm{s}=\left(1+r_\textrm{p}/r_\textrm{s}\right)/\left(r_\textrm{p}+r_\textrm{s}\right)
\]
The fitting procedure is applied two times. First keeping the limb darkening coefficients
fixed at their theoretical values, and afterwards letting them vary. For \textsc{tap} we set
the fitting interval to $\pm0.2$. In case of \textsc{jktebop} we use the option to set the LD
coefficients fixed for the initial model, but let them being perturbed for the error estimation. 
Thus the fitted model does not change, but the error bars are increased.
The eccentricity was fixed to zero for all our analyses.

Since all data are obtained using JD$_\textrm{UTC}$ as time base, we transform the fitted 
mid-transit times to BJD$_\textrm{TDB}$ afterwards using the online converter\footnotemark
\footnotetext{http://astroutils.astronomy.ohio-state.edu/time/utc2bjd.html} provided 
by Jason Eastman 
\citep*[for a detailed description of the barycentric dynamical time see][]{Eastman2010}.

Finally, we derive the photometric noise rate \citep[pnr,][]{Fulton2011} as a quality 
marker for all light curves, which is defined as the ratio between the root mean square 
of the model fit and the number of data points per minute. For further analysis we took
data with $\textrm{pnr}\lesssim4.5$ into account.

\section{Results}
\label{sec:Results}
For every light curve we get six different models, four from \textsc{tap} (for the binned and unbinned 
data with the LD coefficients fixed and free) and two from 
\textsc{jktebop} (for the binned and unbinned data, LD coefficients set free for error estimation only). 
To get one final result we averaged those six results. As for the errors, 
we got four different estimations from \textsc{tap} and 12 from \textsc{jktebop}. 
As final error value, we took the maximum of either the largest of the error estimates, or the spread of the model fit results to use 
a conservative error estimate. It has to be noted, though, that the spread between the different models has always been below size of the error bars.

Binning the light curve threefold using an error weighted mean in principle still 
leaves enough data points during ingress
and egress to be able to fit the transit model to the data while reducing the error bar of an
individual measurement. However, comparing the results of the threefold binned and unbinned data 
we do not see significant differences, neither in the fitted values, nor in the error bars.

The same counts for the differences between the \textsc{tap} models obtained with fixed LD values and those
obtained with the LD coefficients set free to fit. 
For a detailed discussion of the influence
of the LD model on transit light curves see e.g. \citet{Raetz2014}.

\subsection{HAT-P-18b}\label{sec:ResultsH18}
\begin{figure*}
  \includegraphics[width=0.32\textwidth]{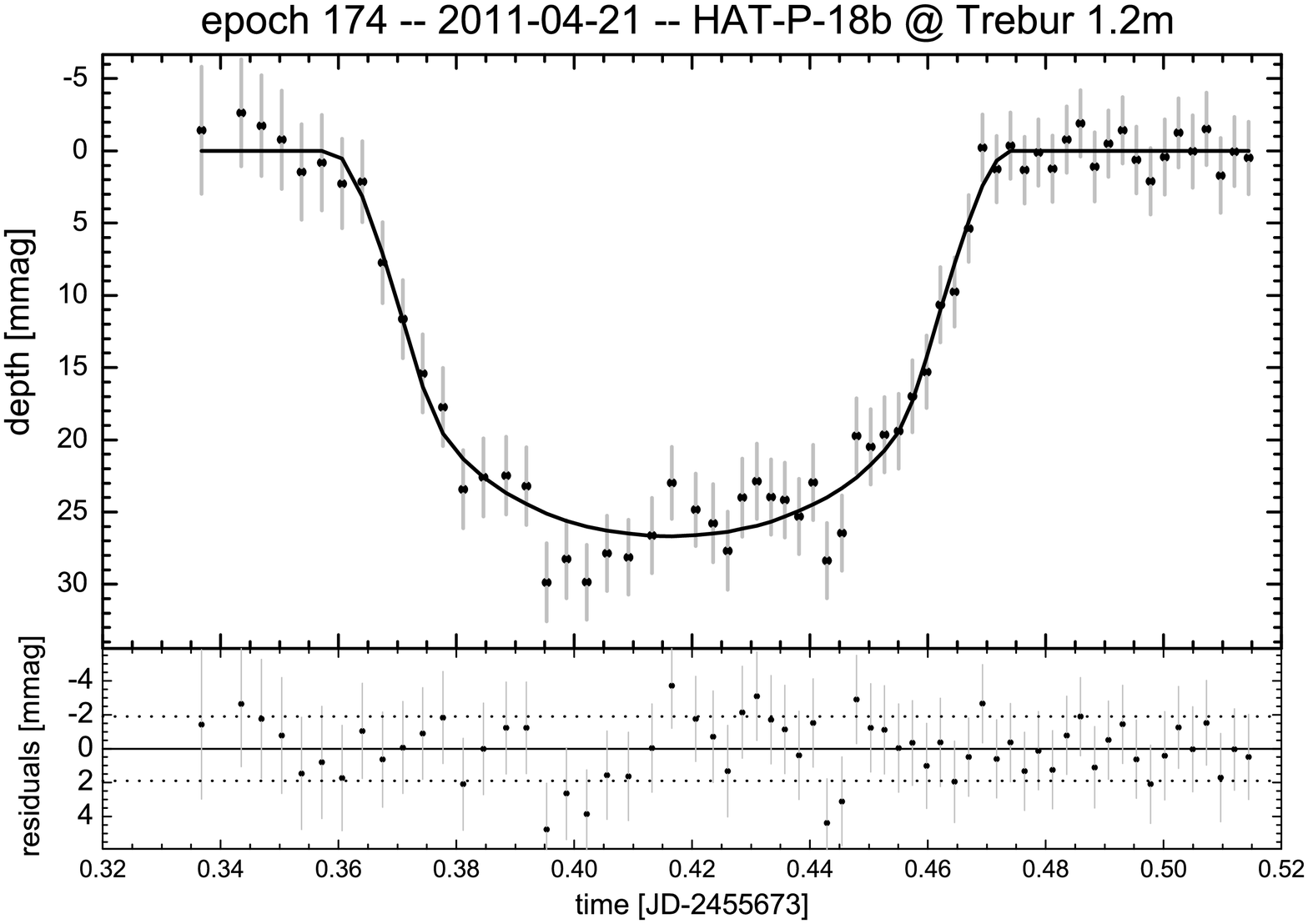}
  \includegraphics[width=0.32\textwidth]{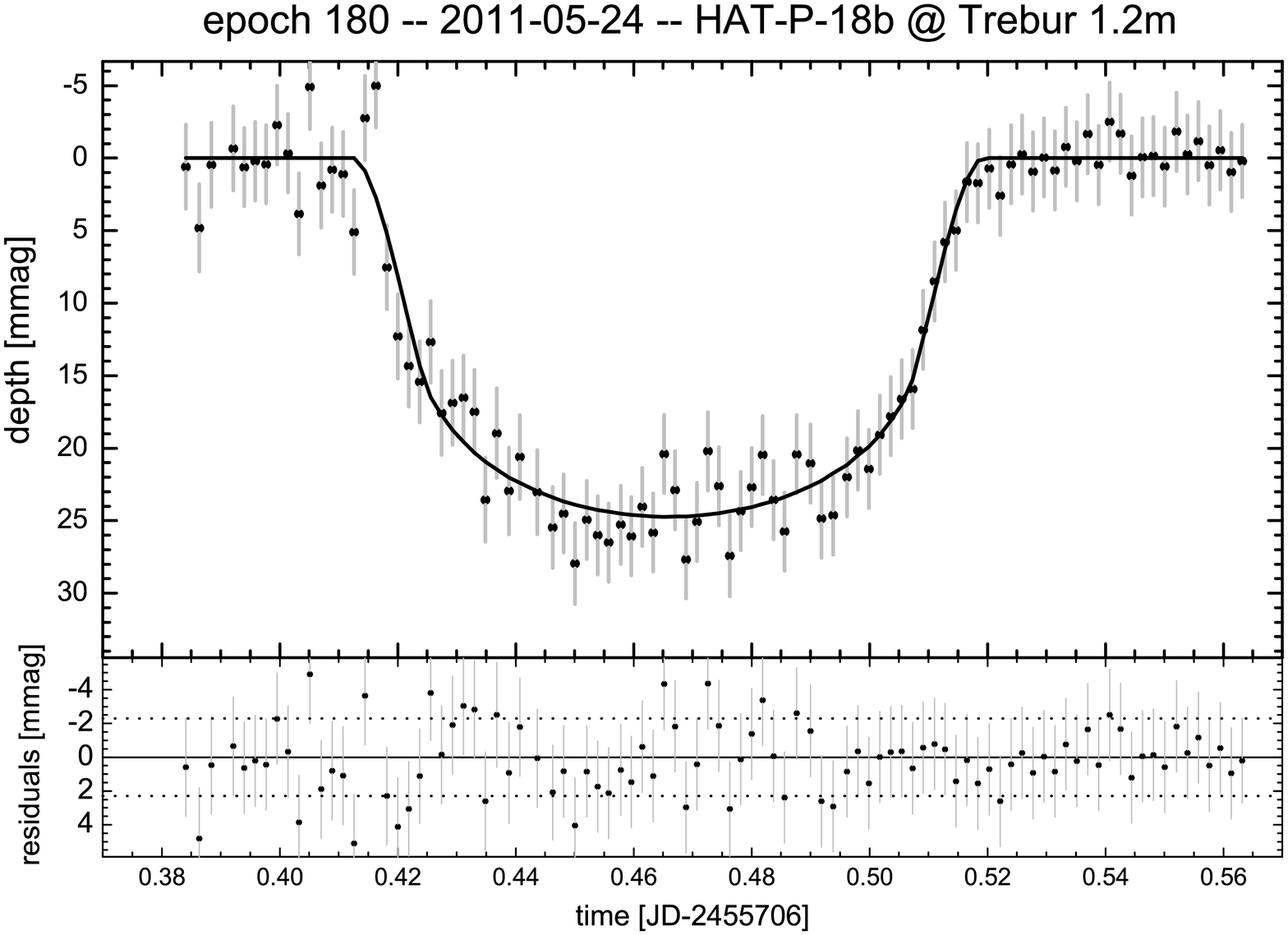}
  \includegraphics[width=0.32\textwidth]{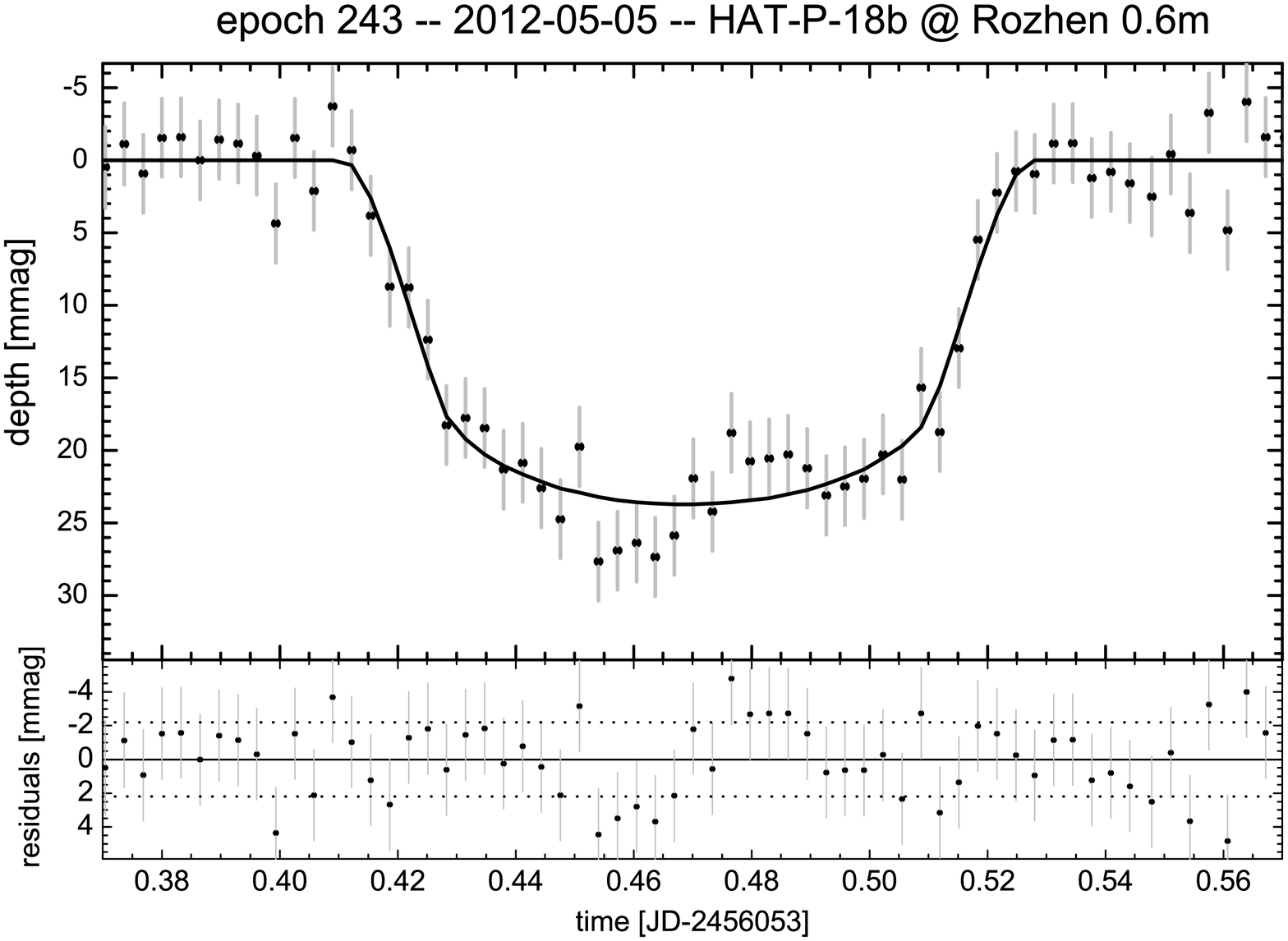}

  %\hrulefill

  \includegraphics[width=0.9\columnwidth]{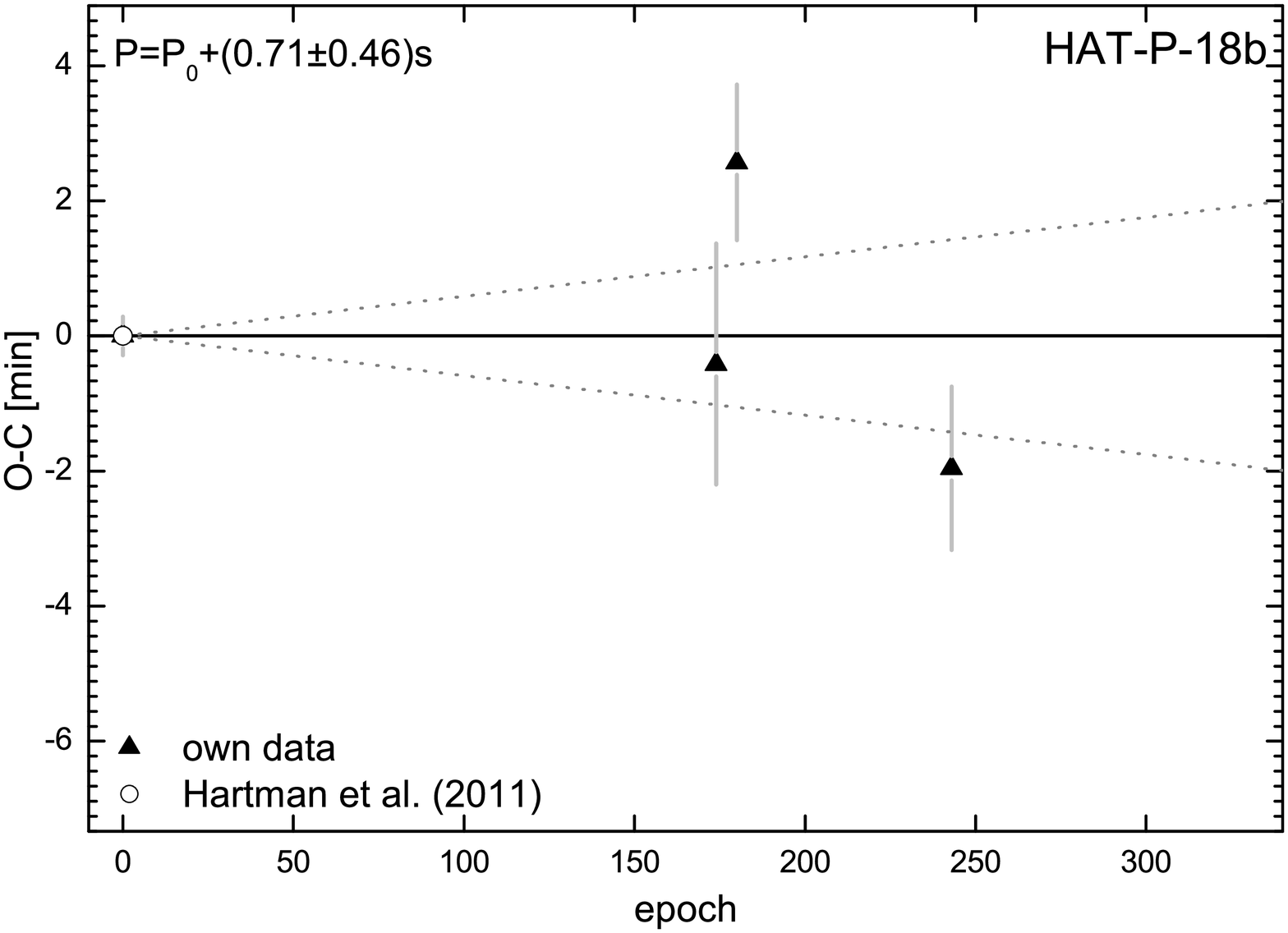}
  \includegraphics[width=0.9\columnwidth]{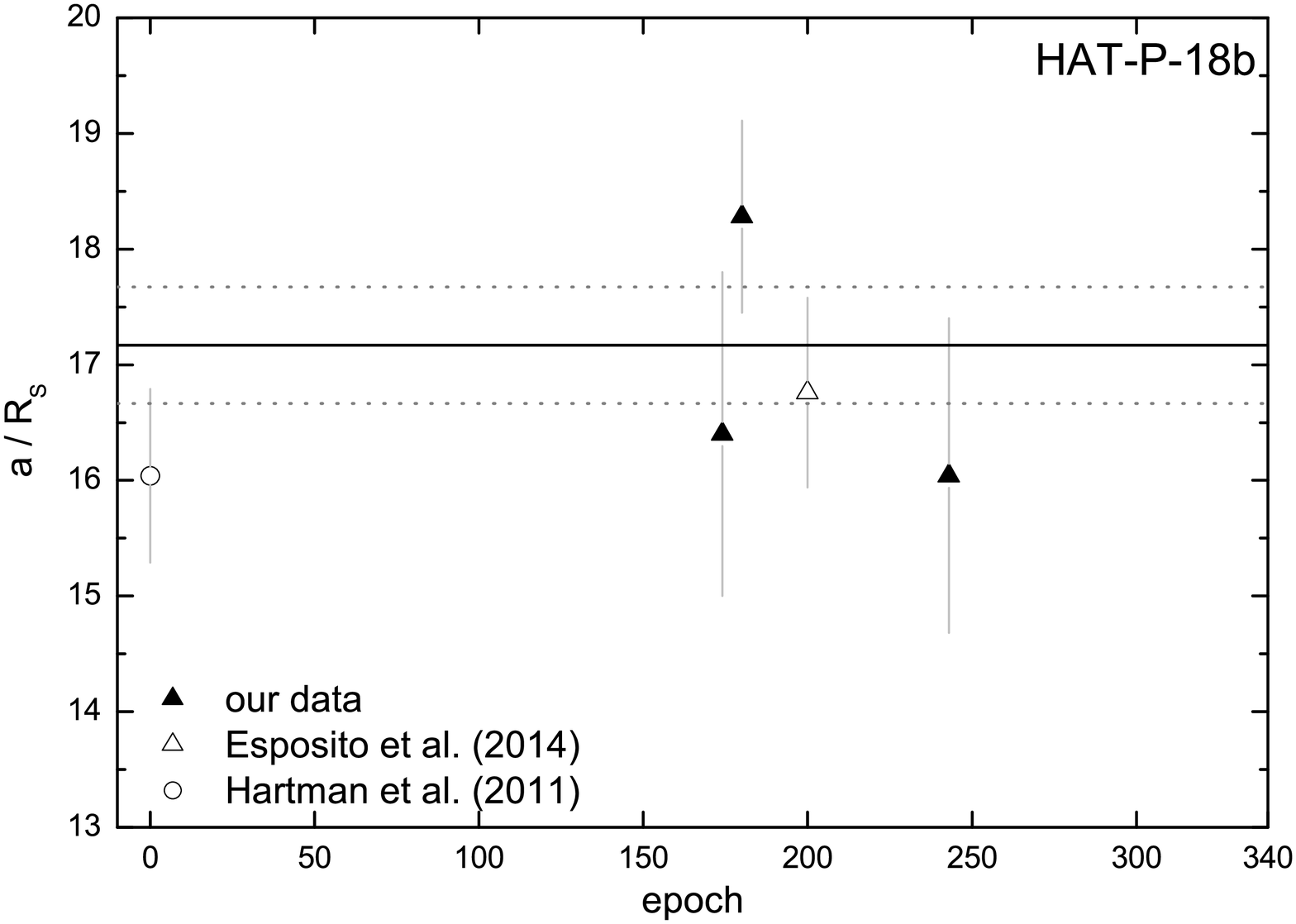}

  \includegraphics[width=0.9\columnwidth]{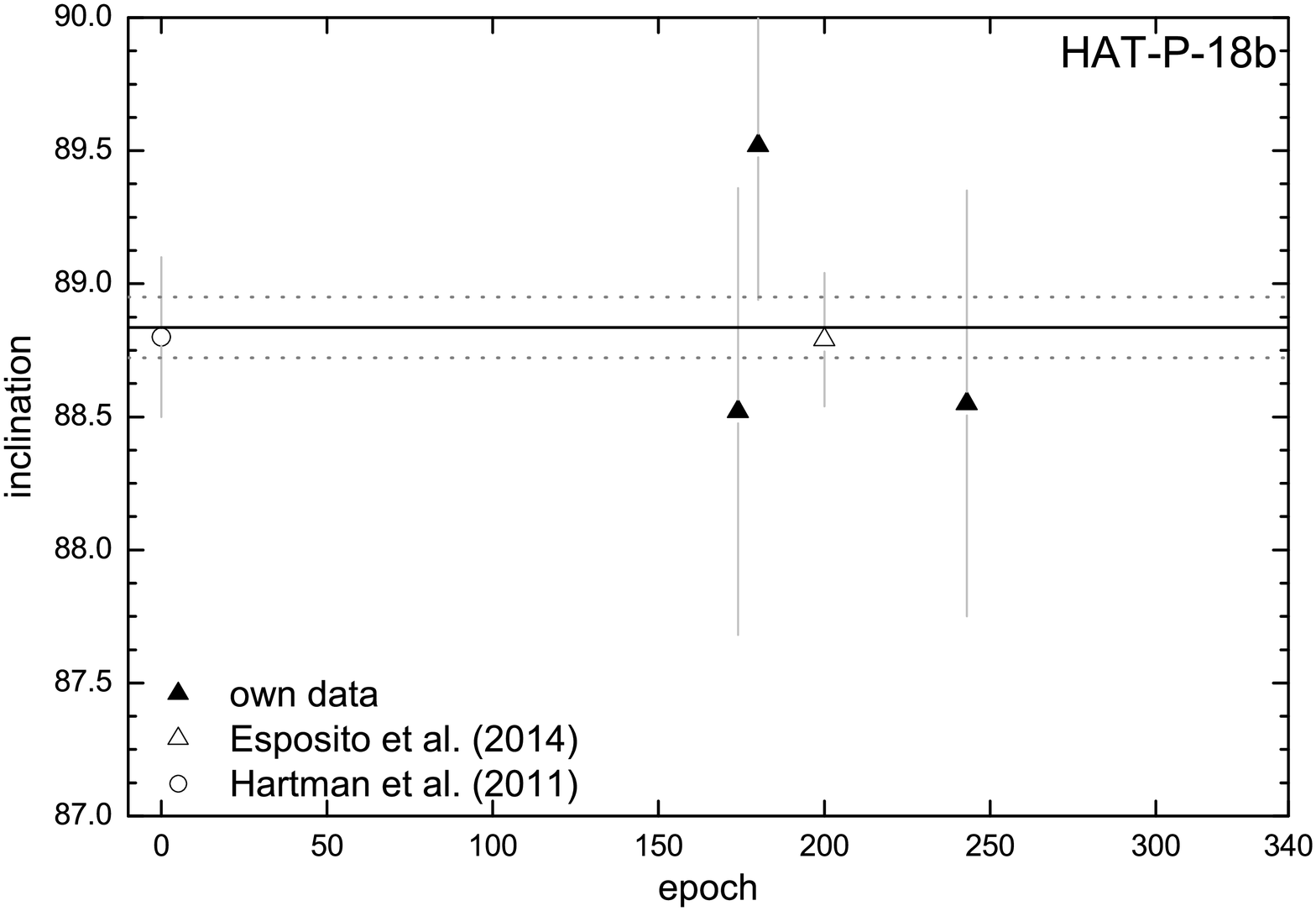}
  \includegraphics[width=0.9\columnwidth]{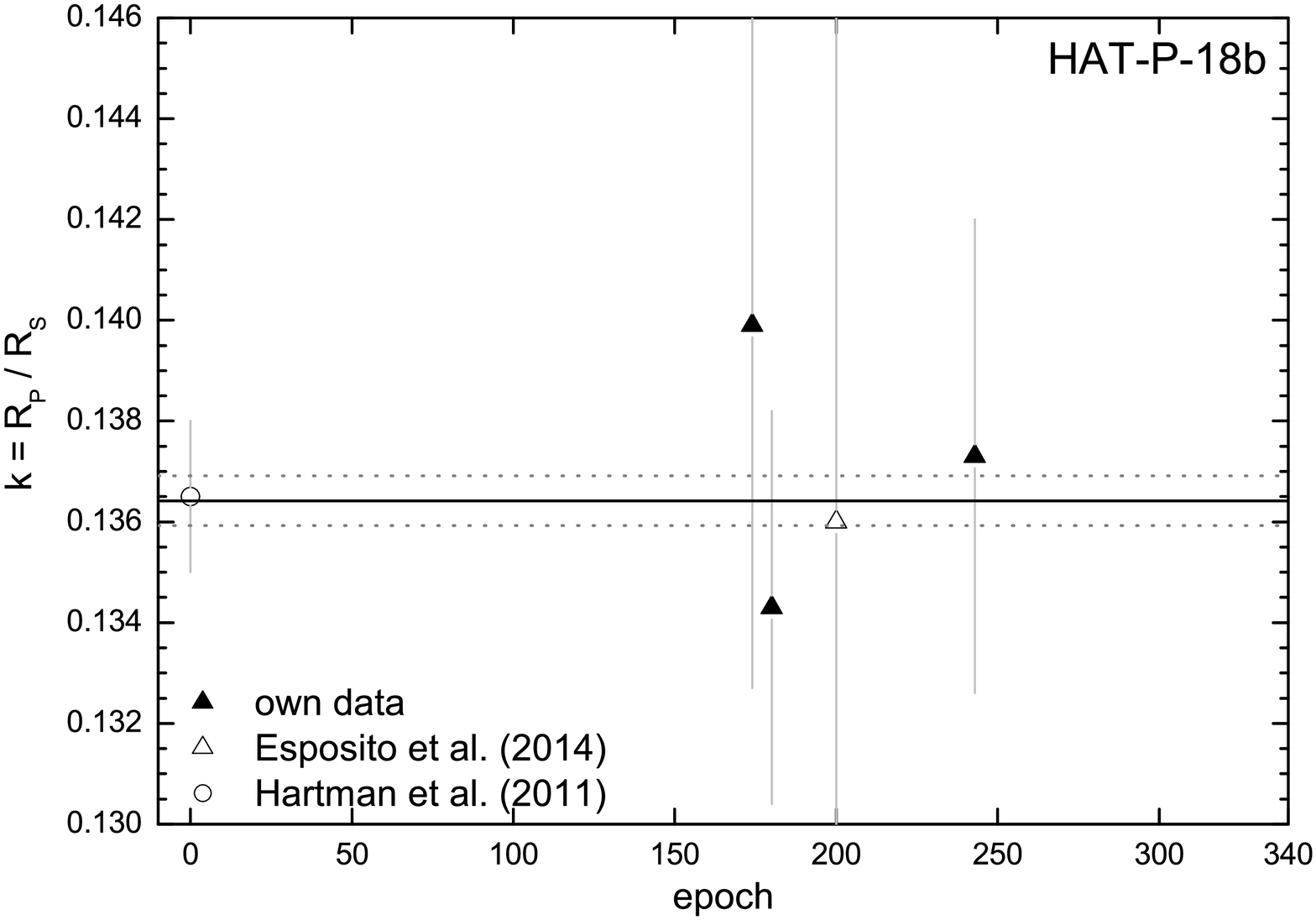}
  \caption{\textit{top:} The threefold binned transit light curves of the three complete transit  
	observations of HAT-P-18b. The upper panels show the light curve, the lower panels 
	show the residuals. The rms of the fit of the threefold 
	binned light curves (dotted lines) are shown as well.
	\textit{bottom:} The present result for the HAT-P-18b observing campaign, including the O--C diagram, 
  as well as the results for the reverse fractional stellar radius $a/R_S$, the inclination $i$, 
  and the planet to star radius ratio $k$. The open circle denotes literature data from
  \citet{HatP18u19}, the open triagles denotes data from \citet{Esposito2014}. Filled triangles denote our data (from Trebur and Rozhen. 
  The dotted line shows the $1\sigma$ error bar of the constant fit.}
  \label{fig:H18_erg}
\end{figure*}

For HAT-P-18b we obtained five transit light curves (see 
Fig~\ref{fig:H18_erg}). All light curves show some features which could be 
caused by stellar activity, e.g. spots. However, the quality of the data does not allow 
to draw further conclusion. Moreover, there is only a small number
of suitable comparison stars available in the respective FoV of each observation, hence this
could also be an artificial effect.

\begin{figure}
  \includegraphics[width=1\columnwidth]{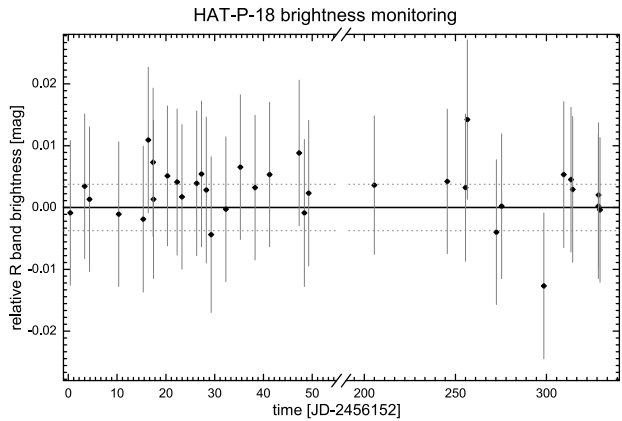}
  \caption{Relative R band brightness of the star HAT-P-18 over a time span of 12 months. The dotted
  line represents the rms of a constant fit.
  }\label{fig:Monitoring_H18} 
\end{figure}

Except for one -- but not significant -- outlier the differences in the 
O--C diagram (see Fig.~\ref{fig:H18_erg}) can be explained by redetermining 
the published period by $\left(0.53\pm0.36\right)\:$s. Hence we find a slightly larger period
compared to the originally proposed period of \citet{HatP18u19}.

Regarding the transit depth, and thus the planet to star radius ratio, we do not see any changes.
However, having a look at the data provided in the
exoplanet transit database \citep[ETD;][]{Podanny2010} one can see that the values for the transit depth 
reported there vary by several
tens of mmag. Such transit depth variations can be caused by close variable stellar
companions placed within the aperture due to the pixel scale of our detectors and
the defocussing of the telescopes. 
Thus we took a more detailed look at the long term variability of the 
parent star. Over a timespan of 12 months we obtained 3 images in four bands (B, V, R, I) in each
clear night using the Jena 0.6m telescope. 
As shown in Fig.~\ref{fig:Monitoring_H18}, the mean variation of the R band 
brightness is $\approx3.8\:$mmag taking the individual error bar of the
measurements into account. However, this variation is too small to be responsible for
the seen transit depth variations. The monitoring of the remaining bands is not shown but leads
to a similar result.
Having a closer look at the light curves listed for
HAT-P-18b in ETD, one can see that a large number is of lower quality. This is especially
true for those light curves responsible for the spread in the tabulated transit depth.
Taking only the higher quality data into account the spread is much smaller. Depending 
on the quality cut the variation can even reach the order of the error bars of the measurements.
Thus we believe that the transit depth variation is negligible. 

Despite a spread in the data, which can be explained by the quality of the light curves, 
we do not see any significant difference for $k$, $i$ 
and $a/R_\textrm{s}$ between the respective observations. A summary of all obtained 
parameters can be found in Table~\ref{tab:AllNewProperties} at the end of the paper, as well as a comparison with
literature values.

\subsection{HAT-P-19b}
\begin{figure*}
  \includegraphics[width=0.32\textwidth]{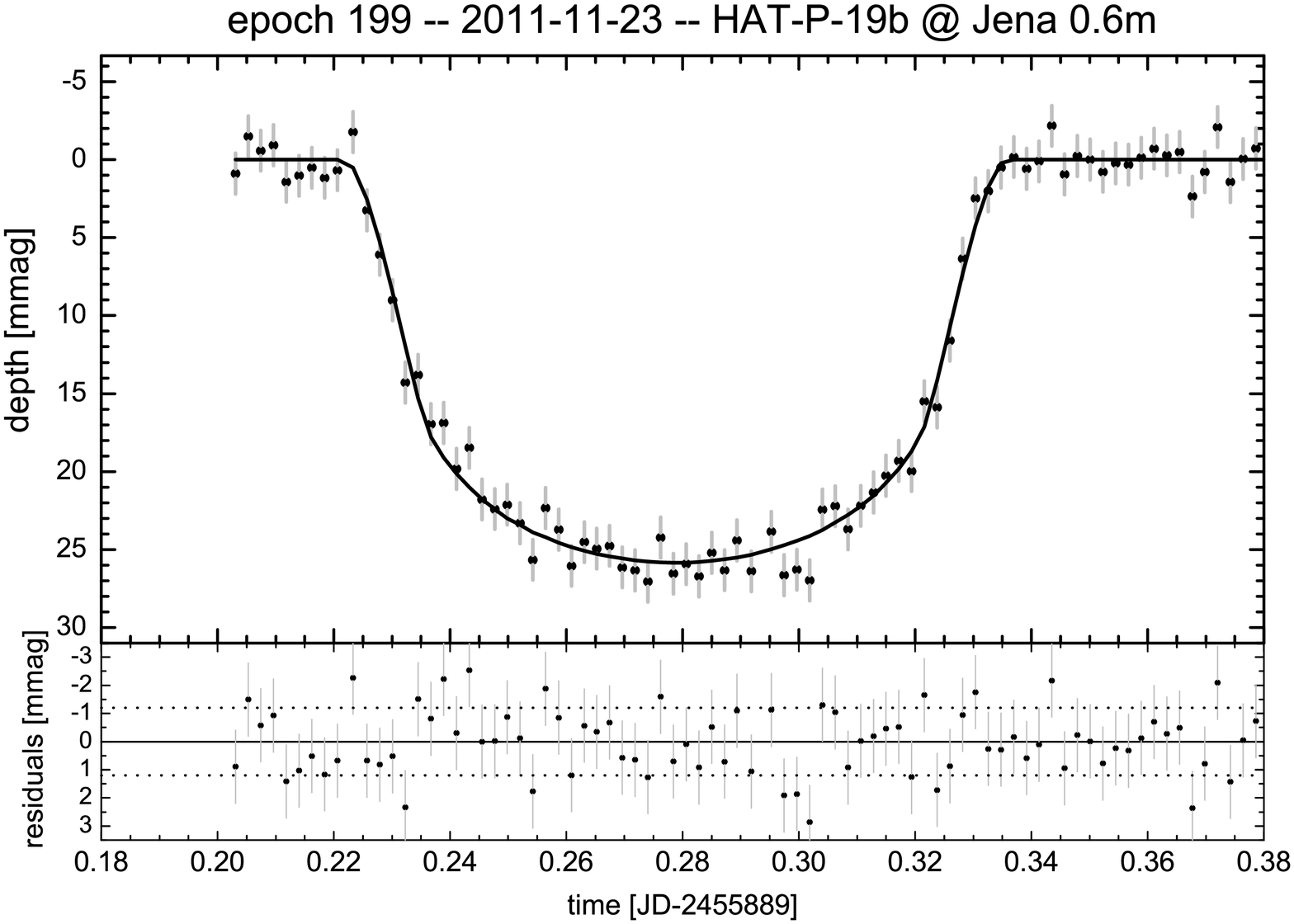}
  \includegraphics[width=0.32\textwidth]{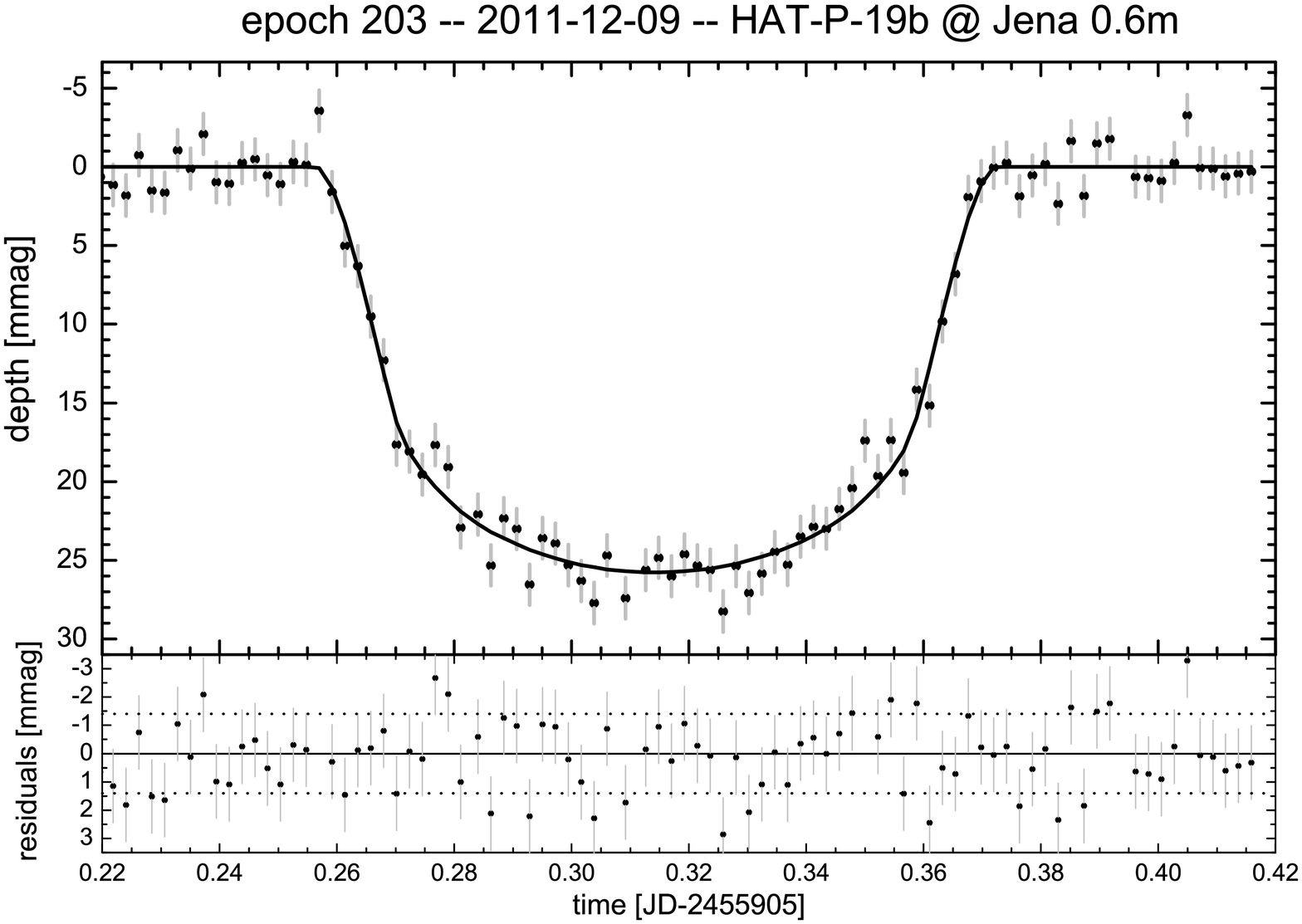}
  \includegraphics[width=0.32\textwidth]{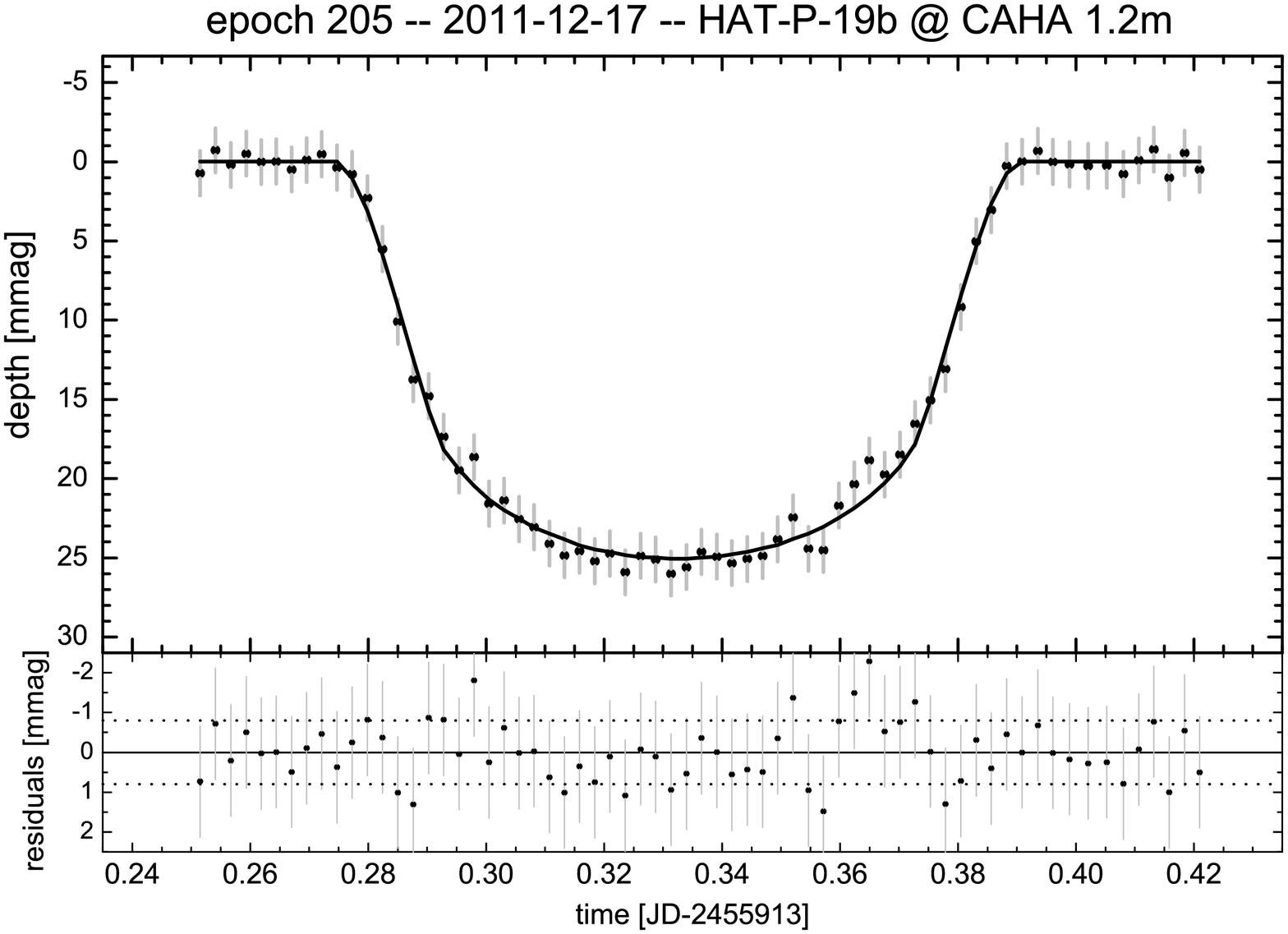}

  \includegraphics[width=0.32\textwidth]{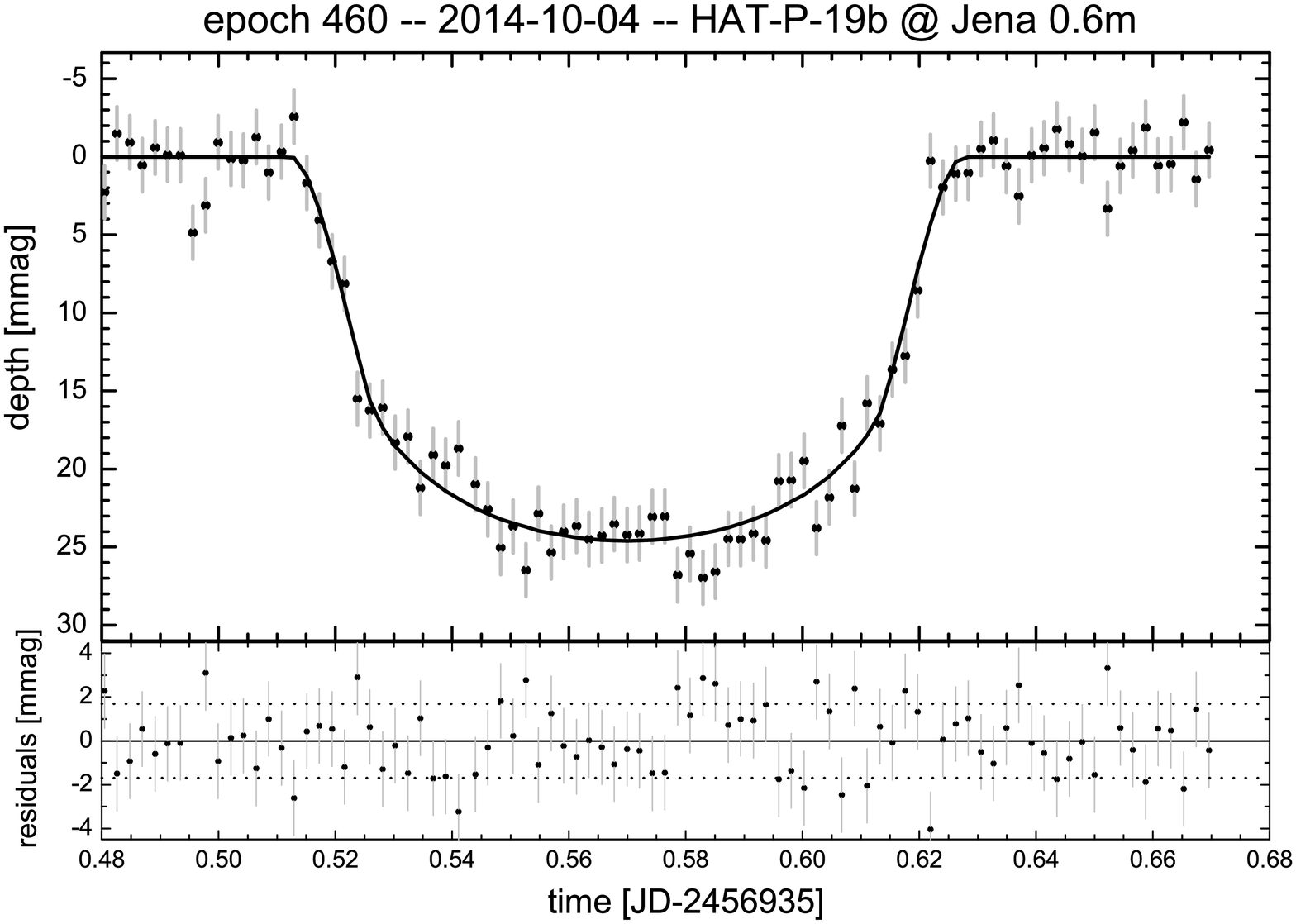}
  %\hrulefill

  \includegraphics[width=0.9\columnwidth]{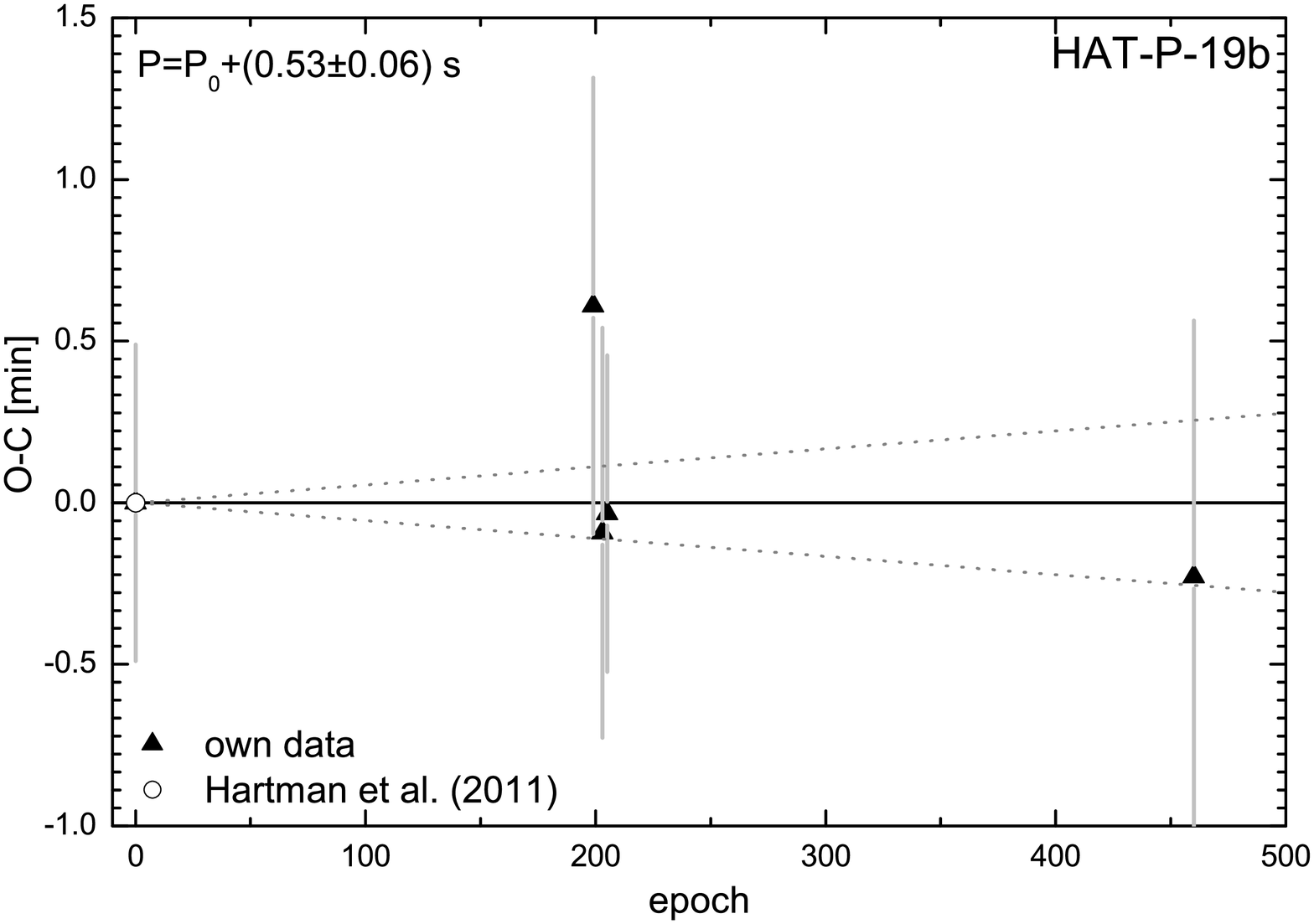}
  \includegraphics[width=0.9\columnwidth]{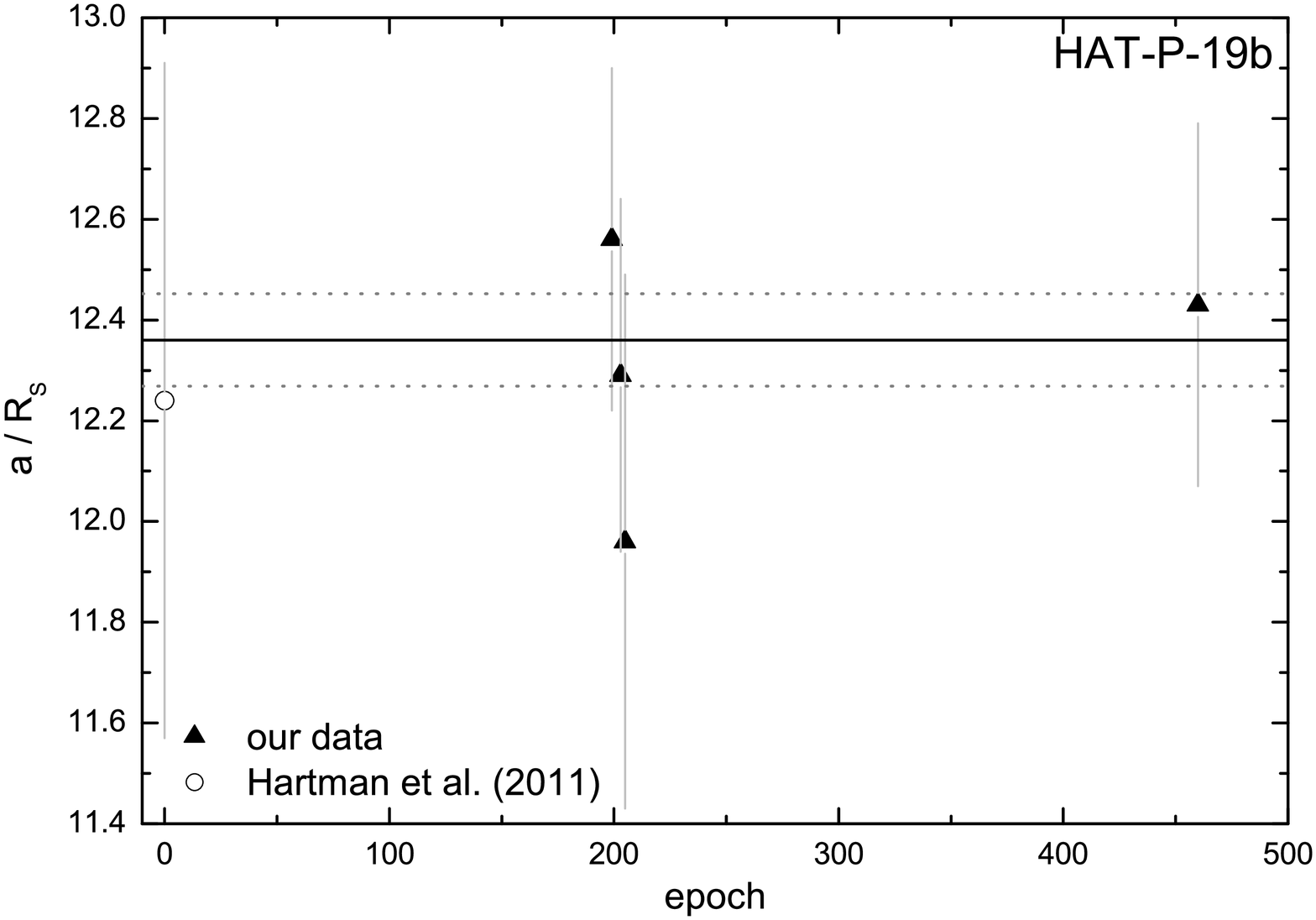}

  \includegraphics[width=0.9\columnwidth]{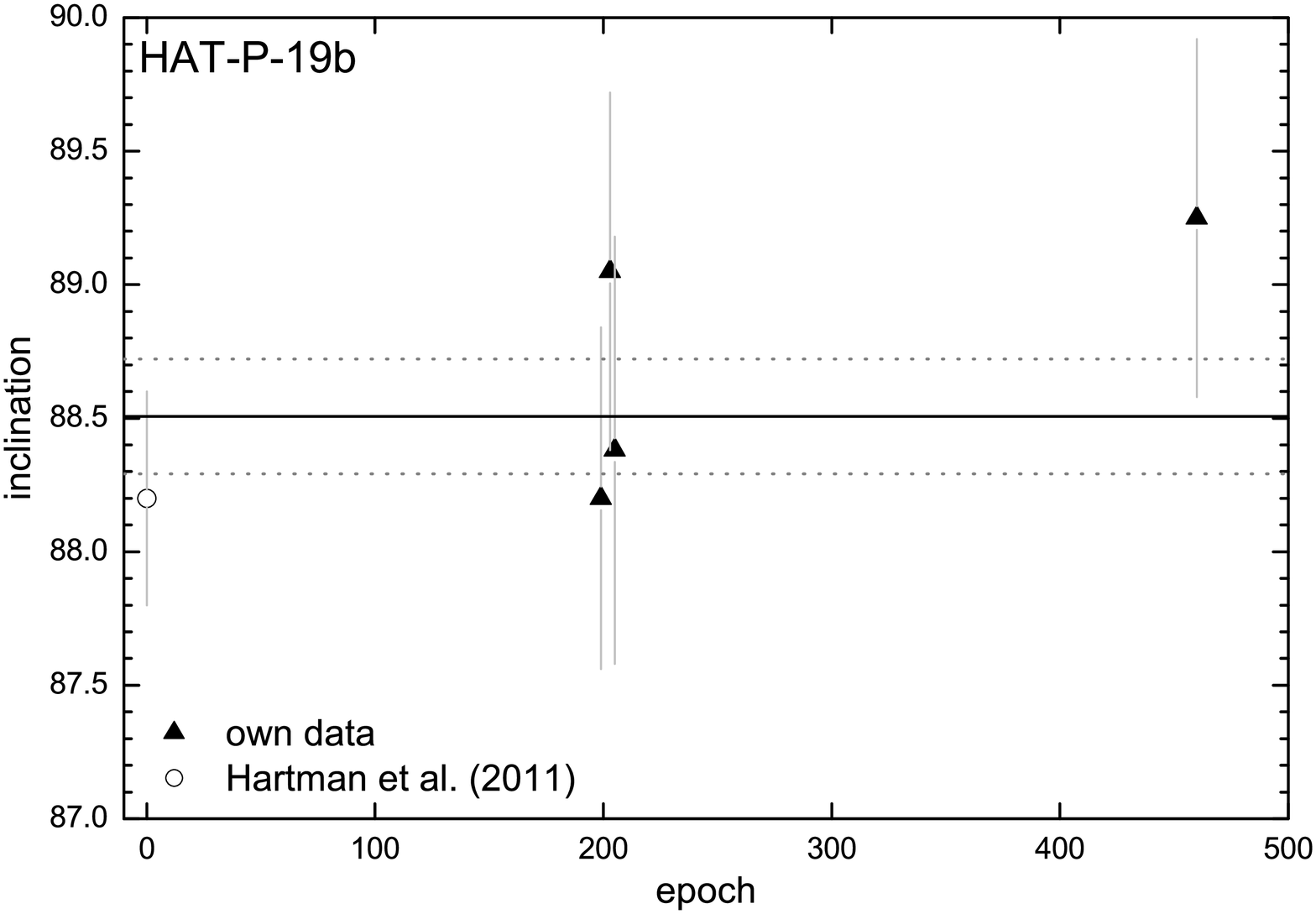}
  \includegraphics[width=0.9\columnwidth]{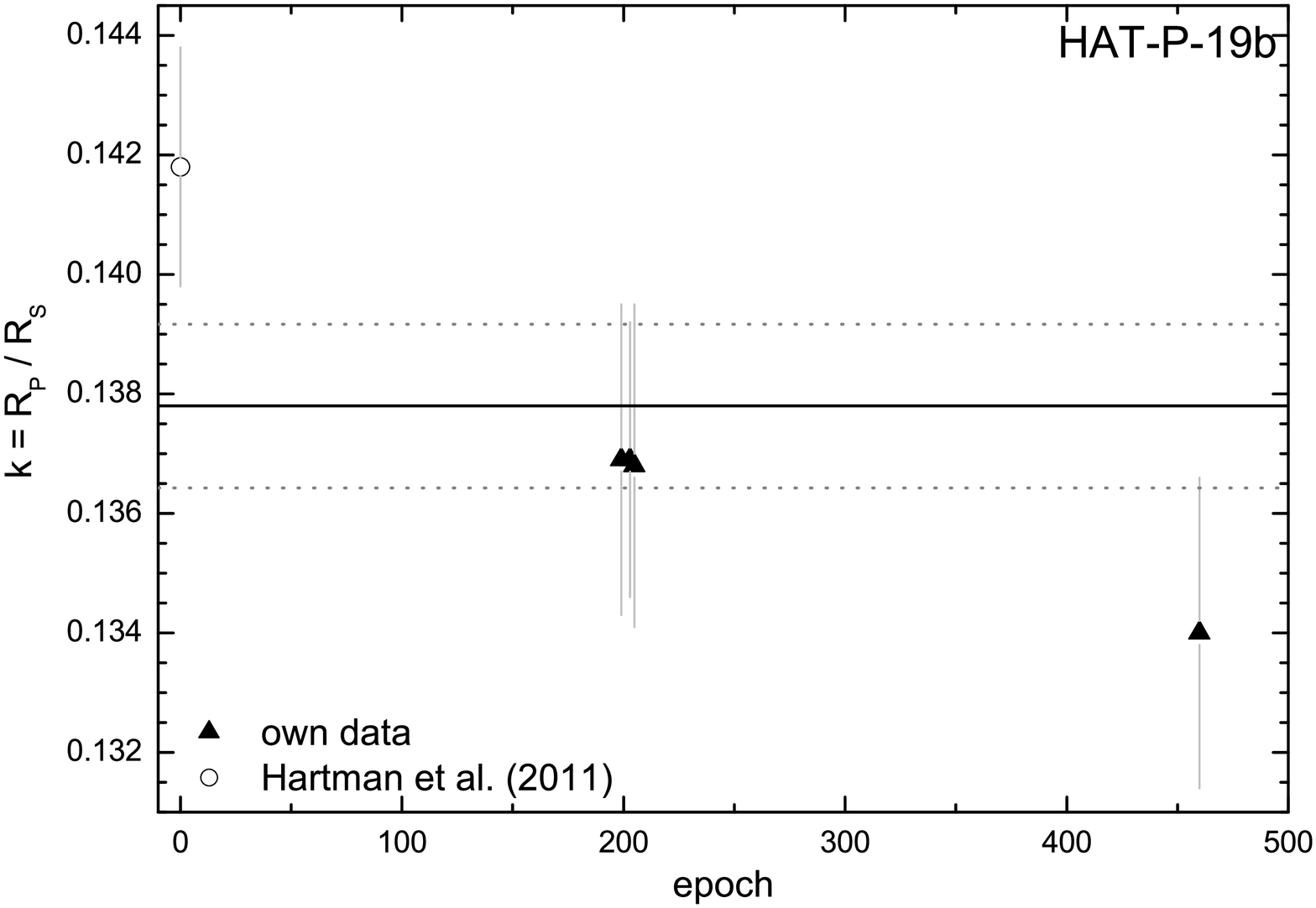}
  \caption{\textit{top:} The transit light curves obtained for HAT-P-19b. \textit{bottom:} 
  The present result for the HAT-P-19b observing campaign. All explanations are equal to 
  Fig.~\ref{fig:H18_erg}. The open circle denotes literature data from
  \citet{HatP18u19}, filled triangles denote our data (from Jena and Calar Alto).}
  \label{fig:H19_erg}
\end{figure*}

For HAT-P-19b we got two light curves using the Jena 0.6m and one light curve from the 
CAHA 1.2m telescope (Fig.~\ref{fig:H19_erg}).
In all three cases we obtained high precision data. The light curves show no
artifacts that could be ascribed to e.g. spots on the stellar surface.
Plotting the mid-transit times into the O--C diagram, we can redetermine the 
period by $\left(0.53\pm0.06\right)\:$s.
As for the inclination and the reverse fractional stellar
radius we can confirm the values reported in \citet{HatP18u19}. The radius ratio of $k=0.1378\pm0.0014$, 
however, seems to be smaller than assumed by \citet{HatP18u19} ($k=0.1418\pm0.0020$).

\subsection{HAT-P-27b}\label{subsec:ergH27}
HAT-P-27b planetary transits were observed six times. An advantage of a network such as
YETI lies within the possibility of simultaneous observations using different 
telescopes. This enables us to independently check whether the data is reliable.
For HAT-P-27b simultaneous observations could be achieved at epoch 415 using two different 
telescopes (Antalya 1.0m and OSN 1.5m).
 
\begin{figure*}
  \includegraphics[width=0.32\textwidth]{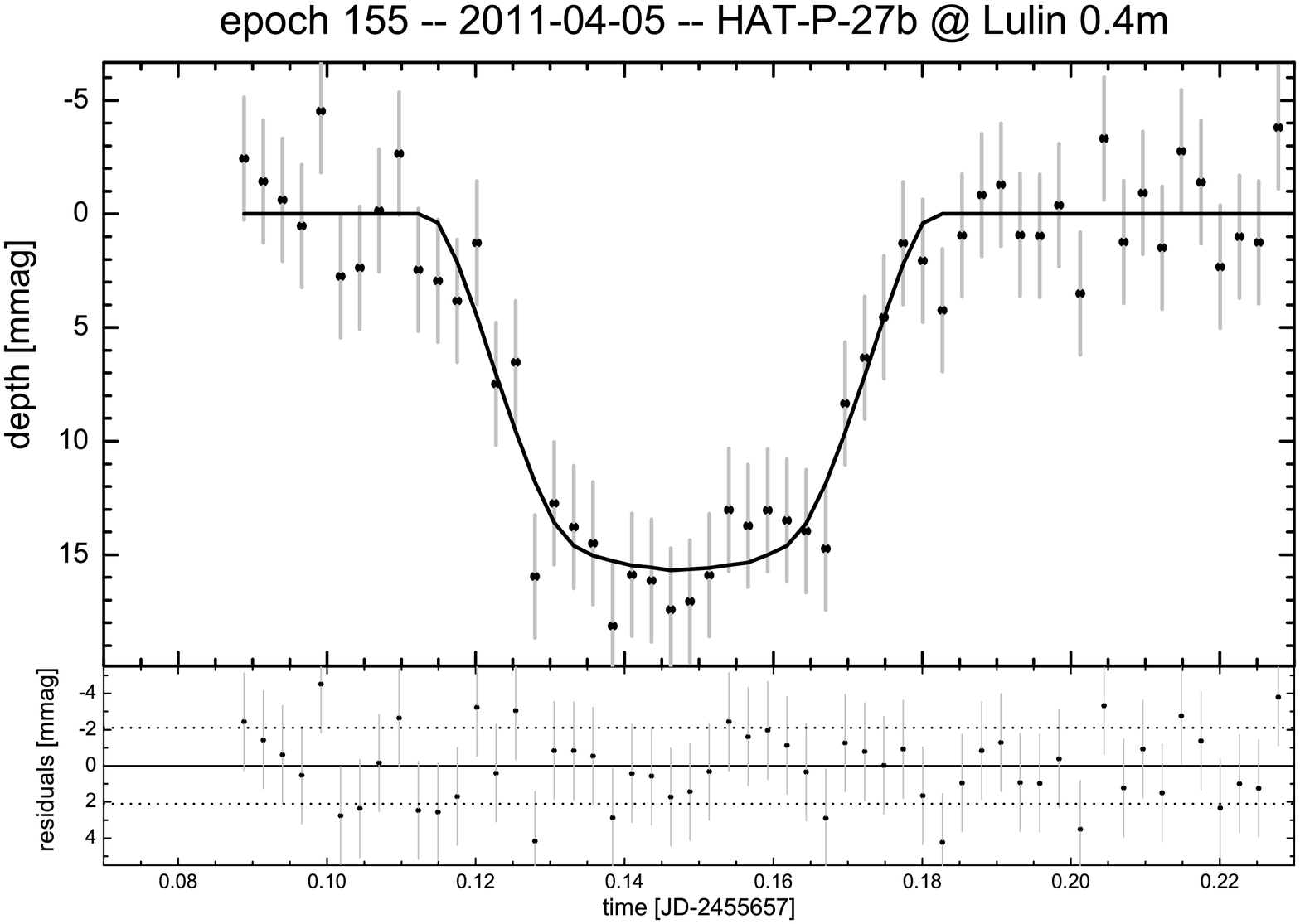}
  \includegraphics[width=0.32\textwidth]{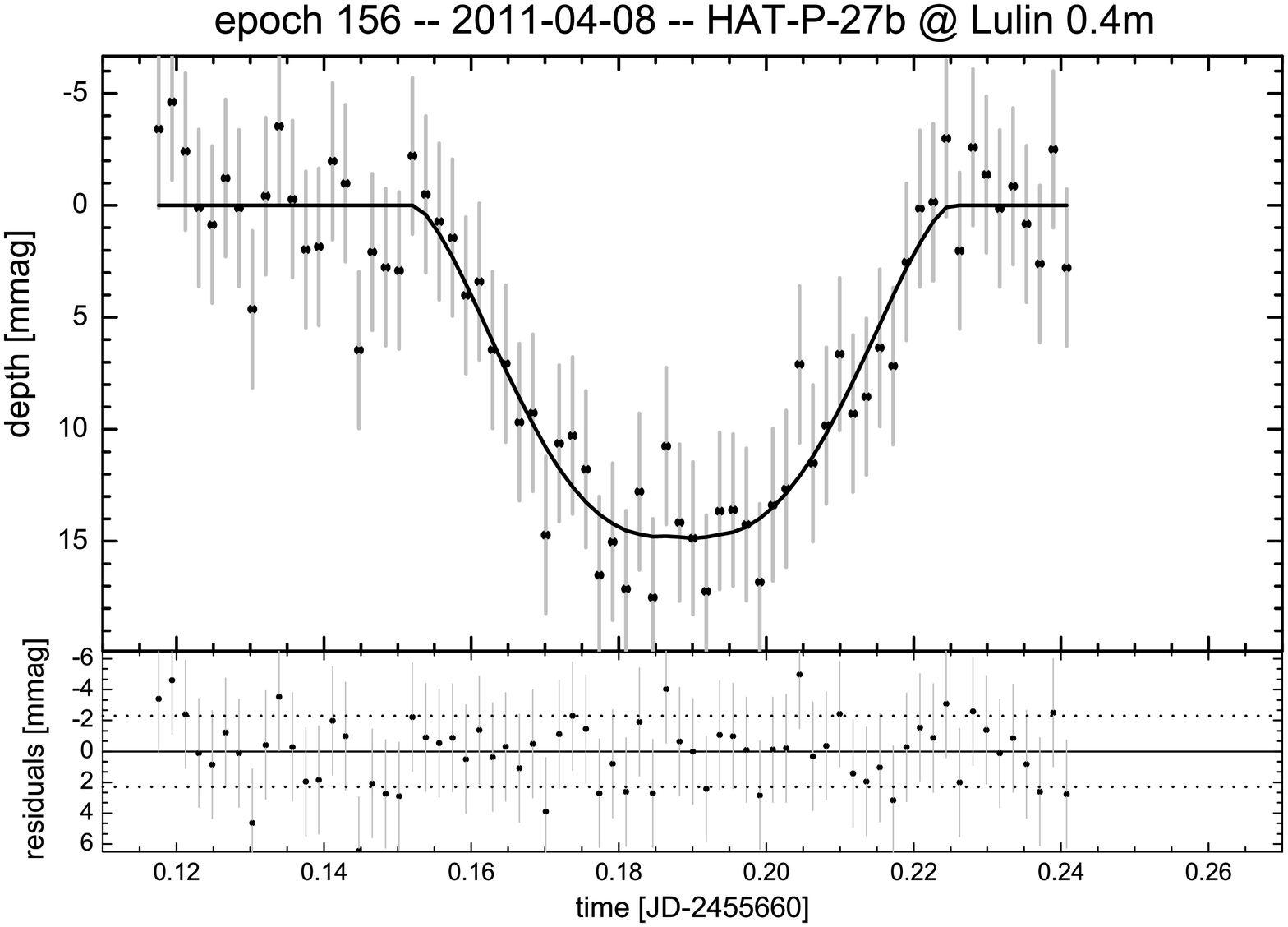}
  \includegraphics[width=0.32\textwidth]{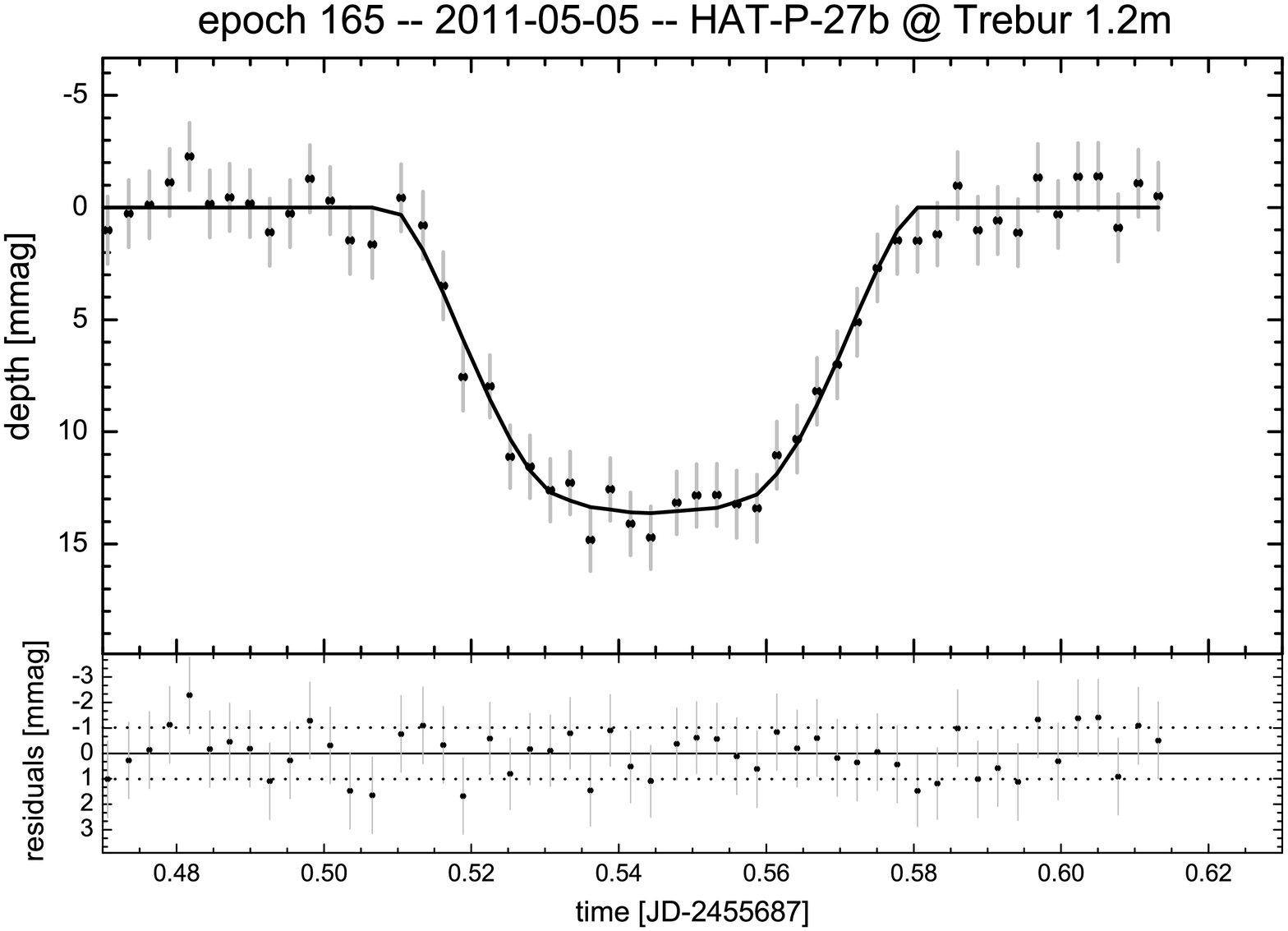}

  \includegraphics[width=0.32\textwidth]{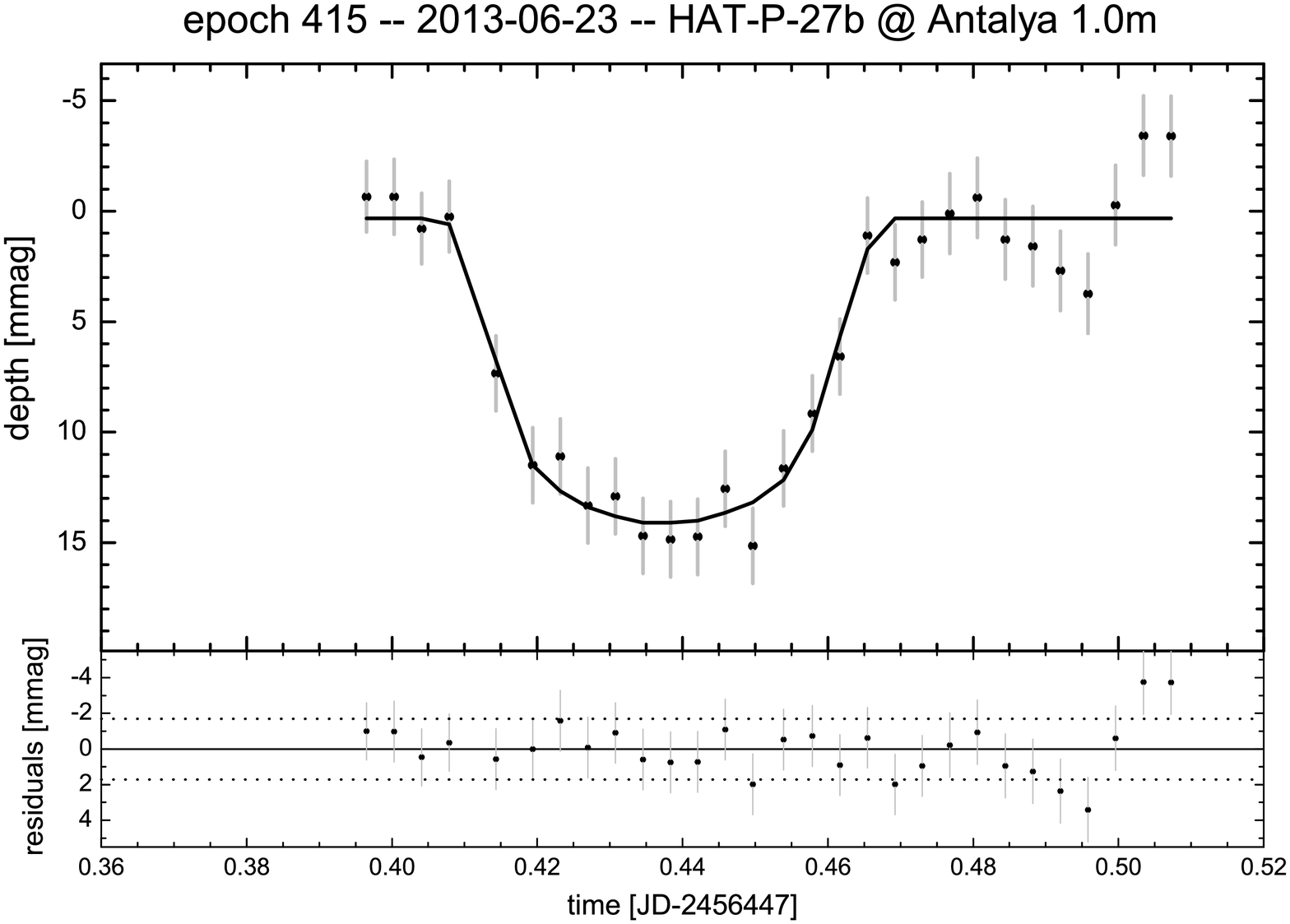}
  \includegraphics[width=0.32\textwidth]{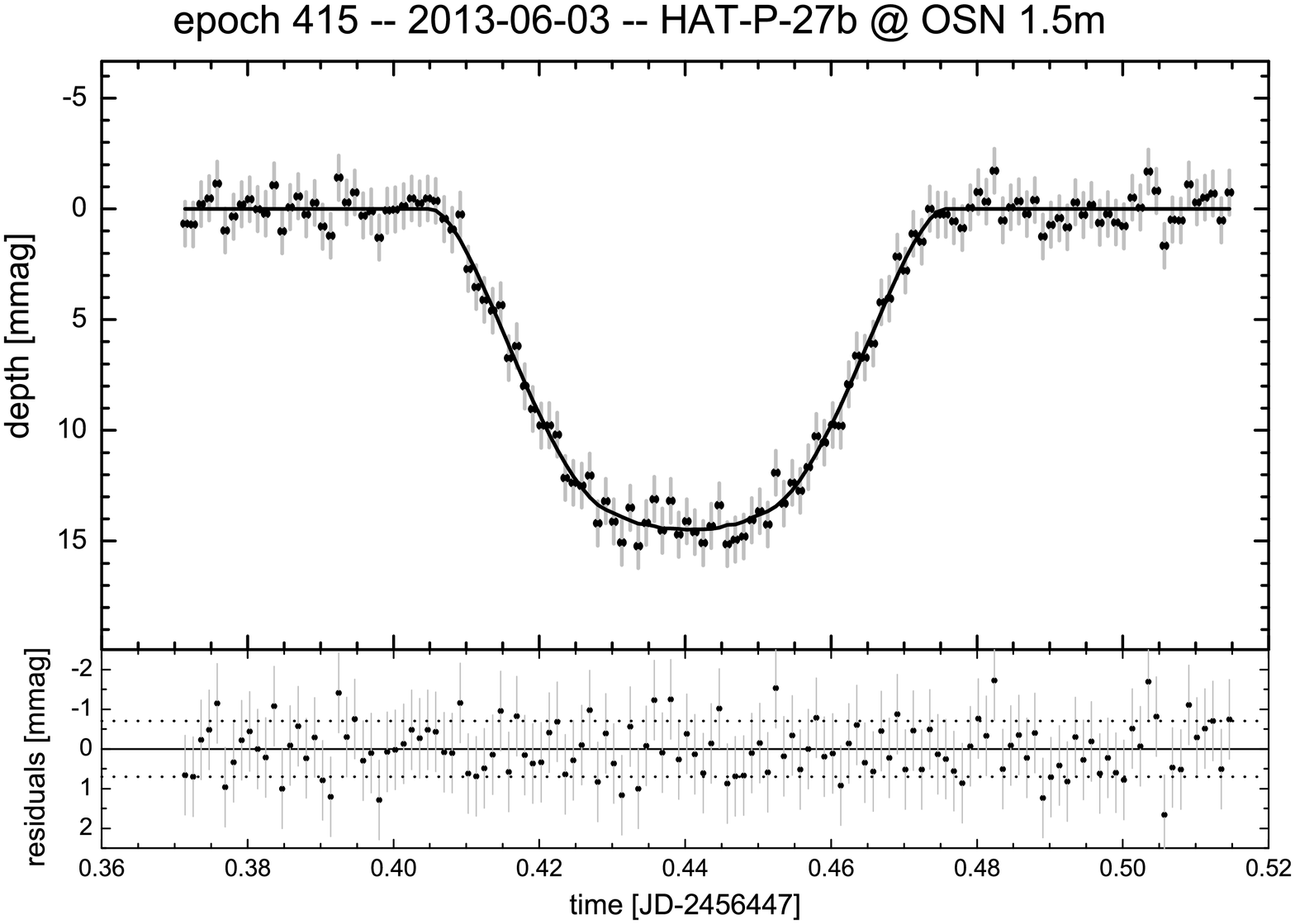}
  \includegraphics[width=0.32\textwidth]{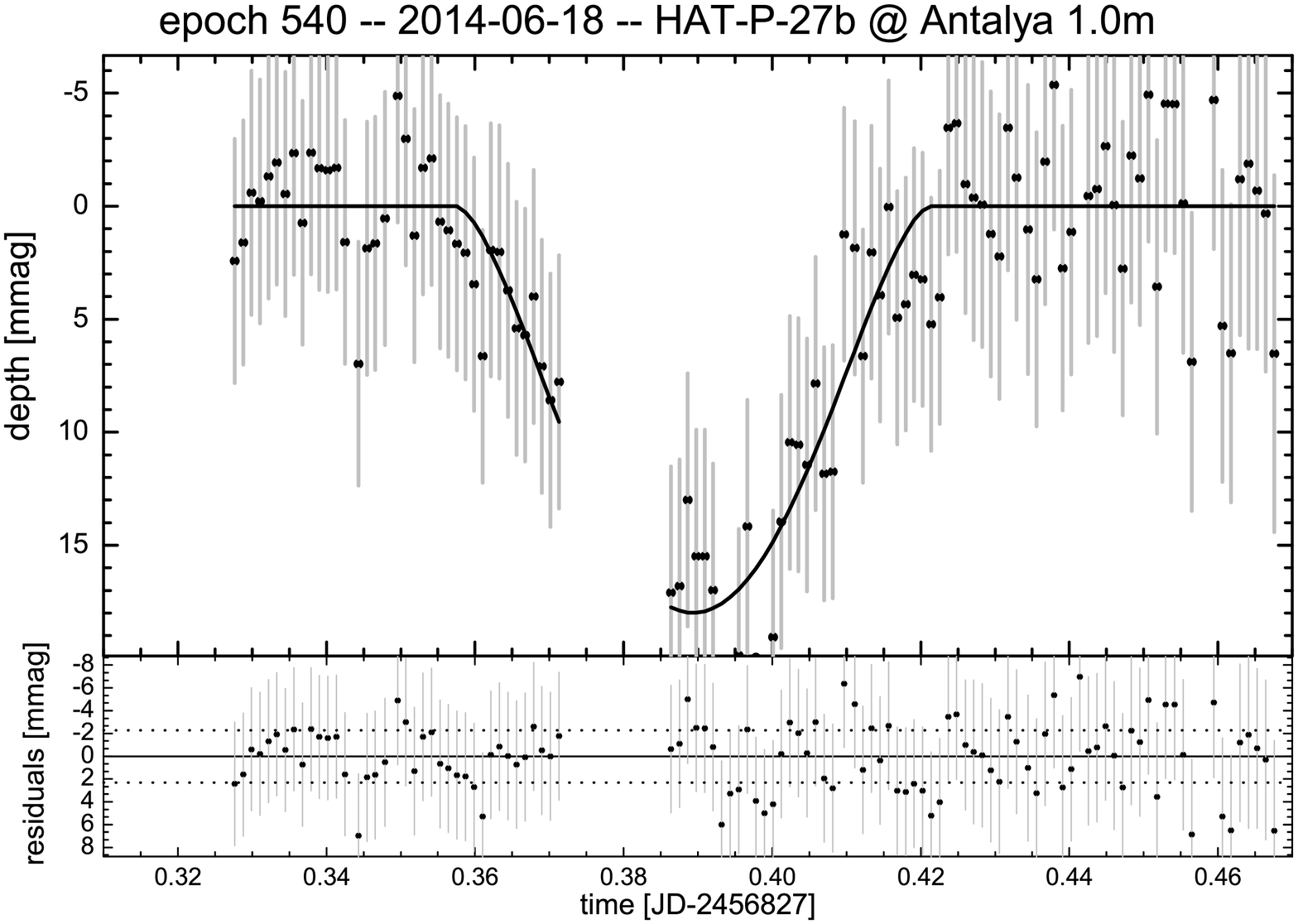}

  %\hrulefill

  \includegraphics[width=0.9\columnwidth]{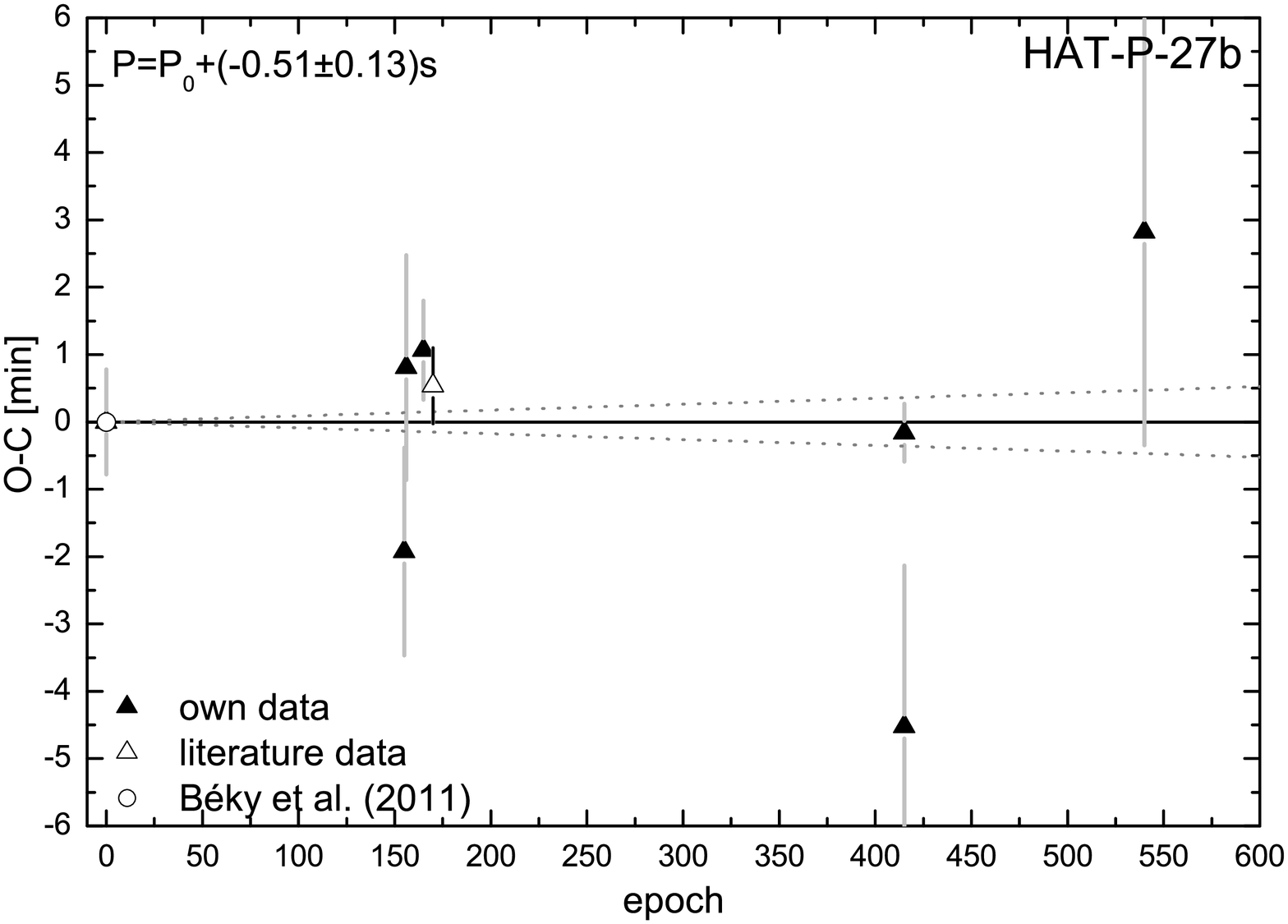}
  \includegraphics[width=0.9\columnwidth]{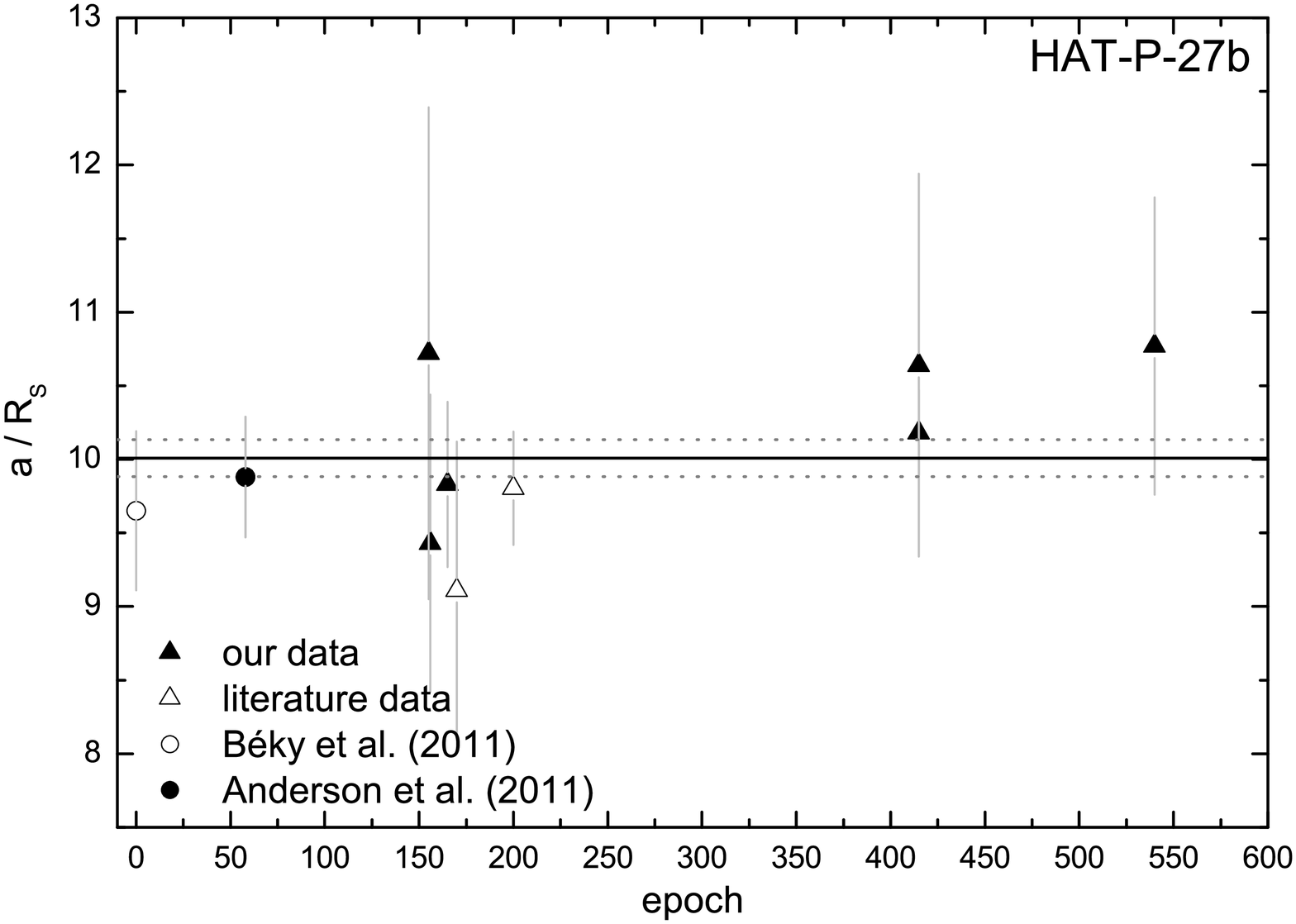}

  \includegraphics[width=0.9\columnwidth]{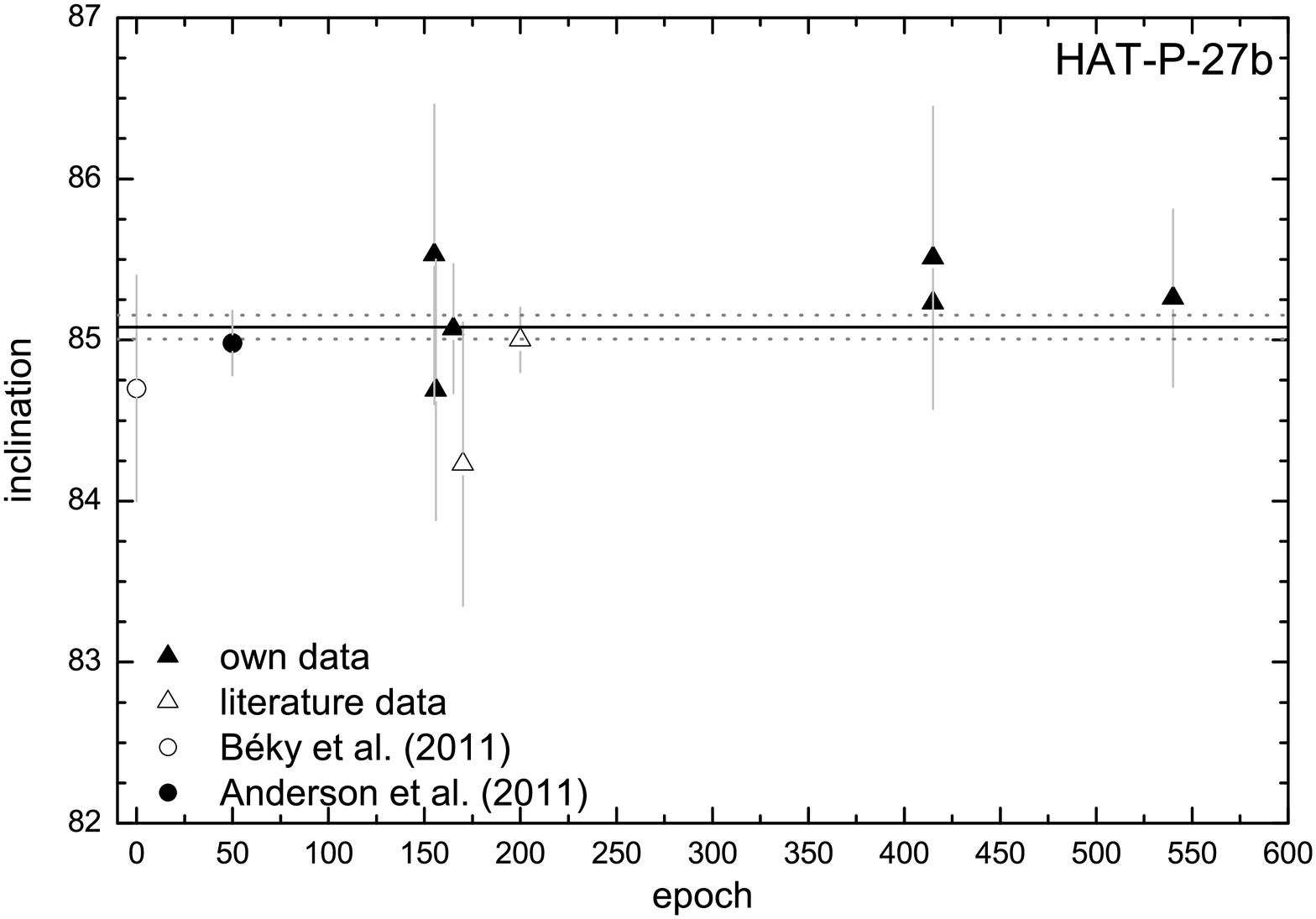}
  \includegraphics[width=0.9\columnwidth]{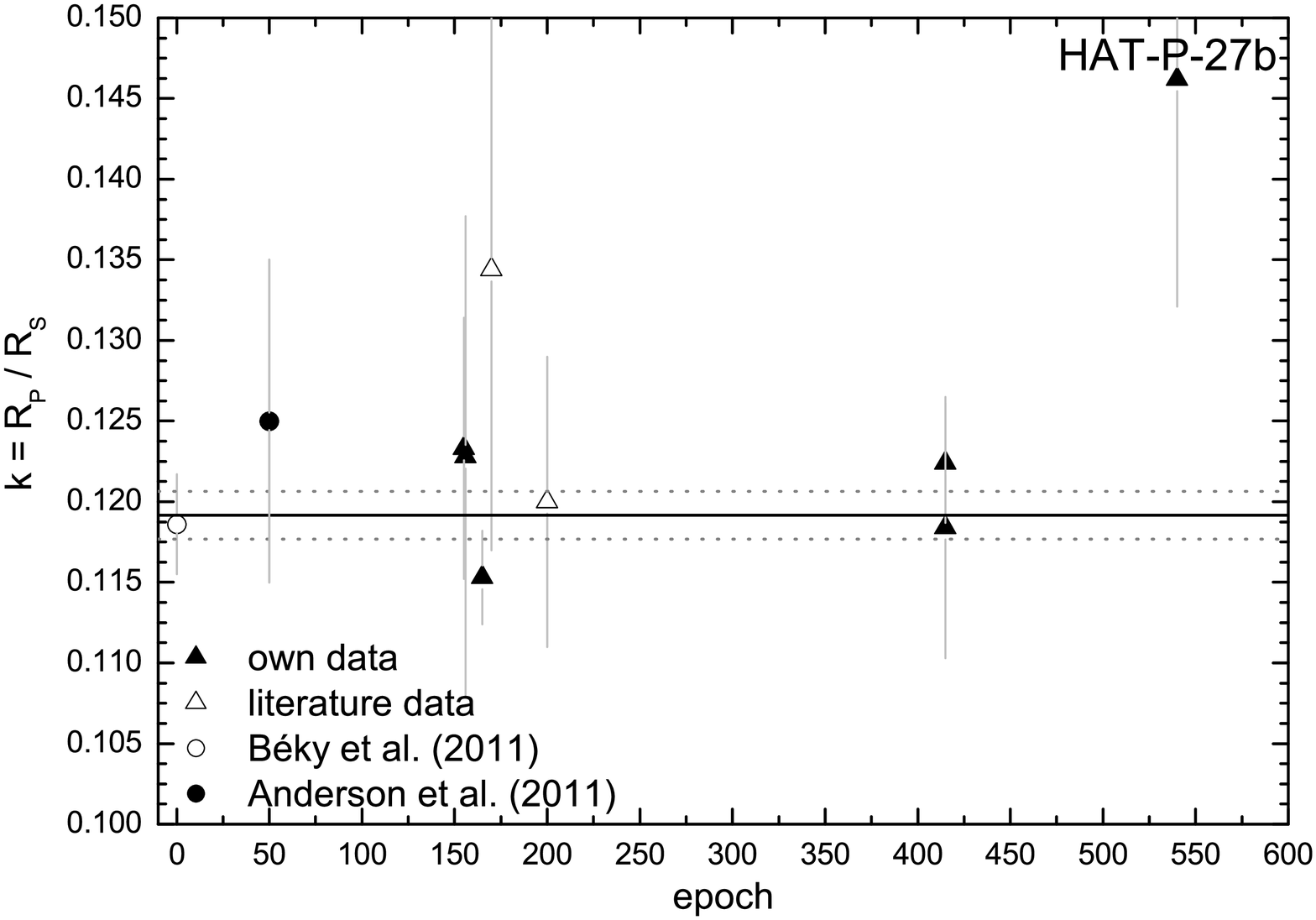}

  \caption{\textit{top:} The transit light curves obtained for HAT-P-27b. \textit{bottom:} 
  The present result for the HAT-P-27b observing campaign. All explanations are equal to 
  Fig.~\ref{fig:H18_erg}. The open circles denotes data from the discovery papers
  of \citet{HatP27} and \citet{Wasp40}, open triangles denote literature data from \citet{Sada} 
  and \citet{Brown2012} (the latter one set to epoch 200 artificially),
  filled triangles denote our data (from Lulin, Trebur, Xinglong and Antalya).}
  \label{fig:H27_erg}
\end{figure*}

As one can see in Fig.~\ref{fig:H27_erg} the best-fitting 
transit shapes differ. Since some data has been acquired using small telescopes under 
unfavorable conditions, this might be an artificial effect. However, these shape variations
can also be seen in the literature data. While \citet{HatP27} fitted a flat bottom to 
the transit they observed, the transit of \citet{Wasp40} shows a rather roundish bottom 
and so does the \citet{Sada} transit. Both \citet{Wasp40}
and \citet{Sada} claim, a roundish bottom is in good agreement with a grazing transit.
Our most precise transit light curve of HAT-P-27b at epoch 415 the OSN 1.5m
telescope also shows no flat bottom. Thus, we would agree with the previous authors that 
the planet is grazing. The epoch 540 observations also shows a V-shape. However, due to a 
connection problem parts of the ingress data is missing, thus precise
fits of the transit parameters are not possible.

In addition to our own data we added data from \citet{Sada} and \citet{Brown2012}. 
The latter one only lists system parameters without giving an epoch of observation, 
thus we artificially put them to epoch 200. Since we do not see any systematic trend in the
remaining data, this does not effect the conclusion on the system parameters but improves
the precision of the constant fit.

The system parameters $i$, $a/R_S$ and $k$ can be determined more precisely 
than before taking the errors of the individual measurements into account. 
All three parameters are in good agreement with the results of previous authors.
Furthermore, we do not see any significant variation. The larger $k$-value of the epoch 540 observations
are due to the quality of the corresponding light curve.

Looking at the mid-transit time we see that a period change of 
$\left(-0.51\pm0.12\right)\:$s explains the data quite well. 
The mid-transit time of one of the epoch 415 observations was found to be $\approx 4.5\:$min
ahead of time, while the other one is as predicted. This way we could identify a synchronization
error during one of the observations. This example shows the importance of simultaneous 
transit observations. Unfortunately this was the only sucessful observation of that kind 
within this project \citep[for a larger set of double and threefold observations see e.g.][]{Seeliger2014}.

\subsection{WASP-21b}
\begin{figure*}
  \includegraphics[width=0.32\textwidth]{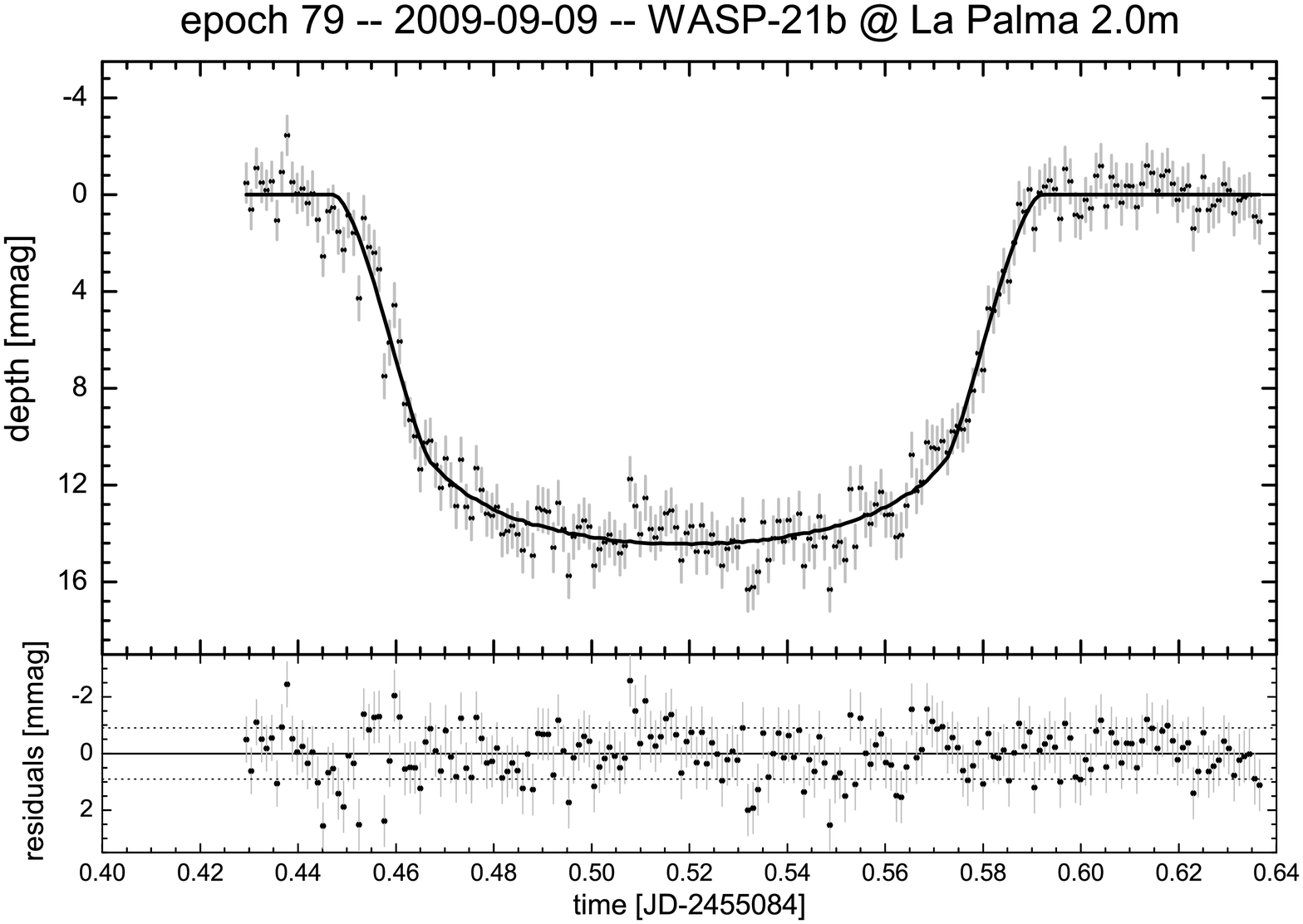}
  \includegraphics[width=0.32\textwidth]{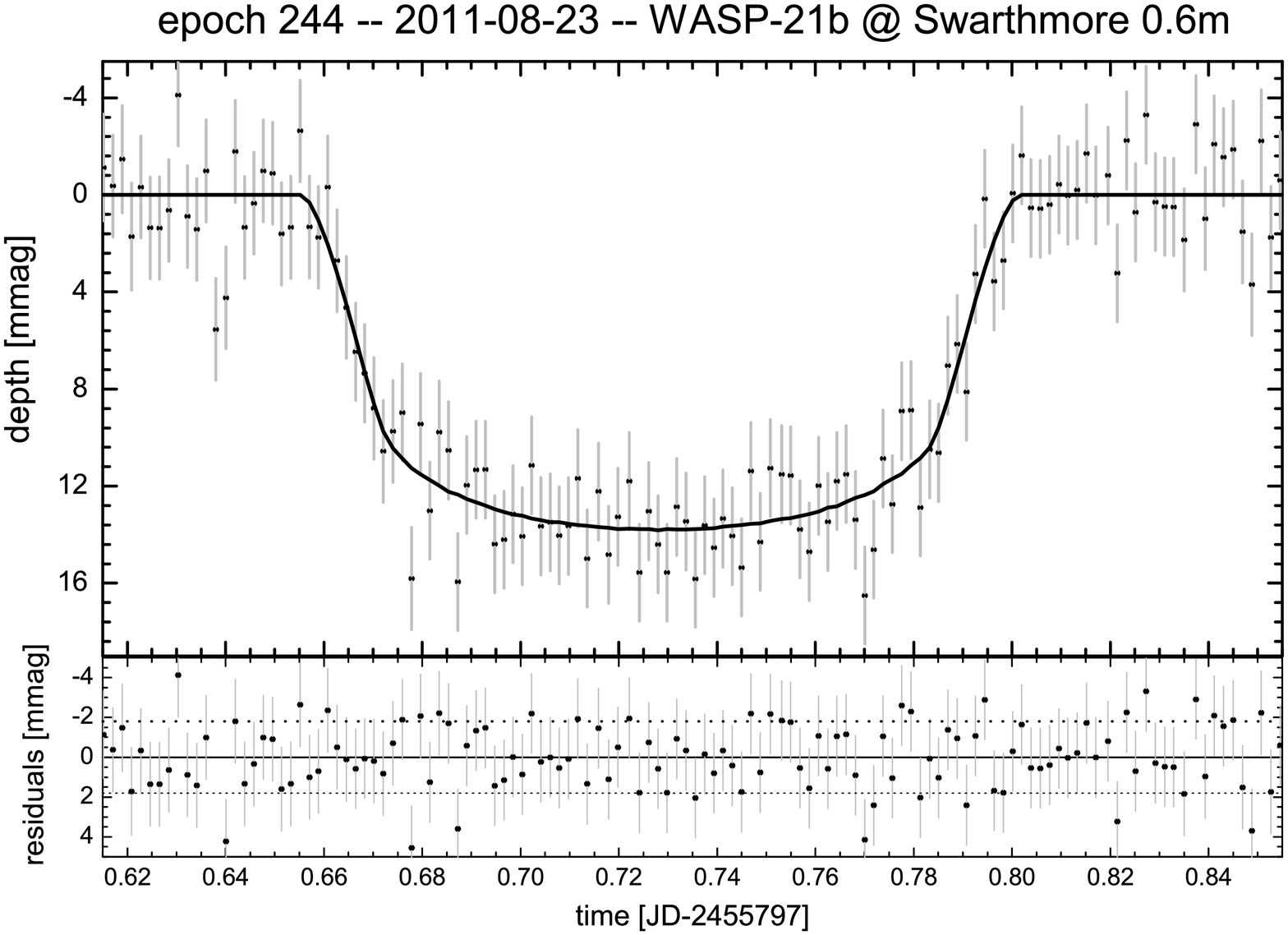}
  \includegraphics[width=0.32\textwidth]{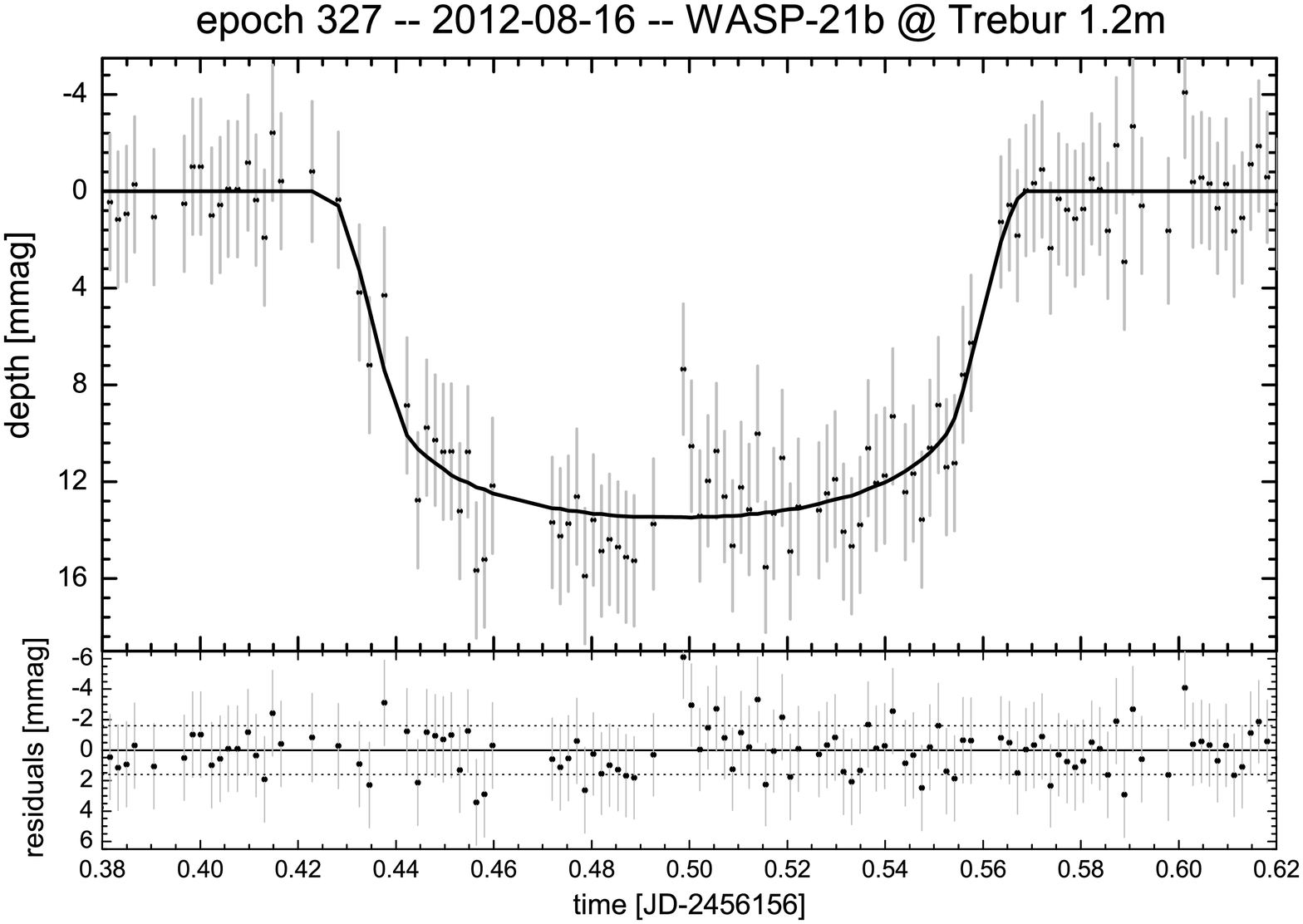}

  \includegraphics[width=0.32\textwidth]{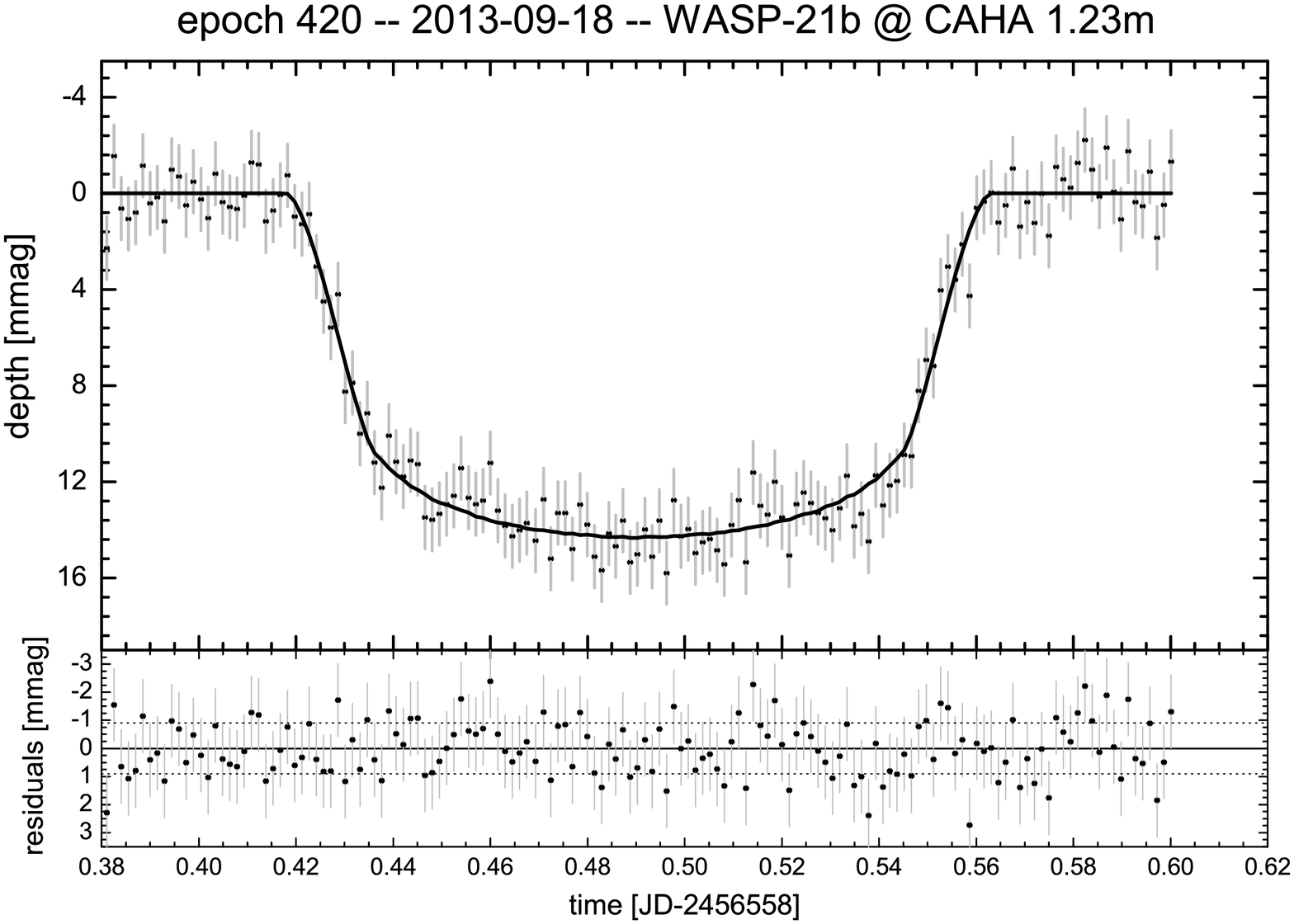}

  %\hrulefill

  \includegraphics[width=0.9\columnwidth]{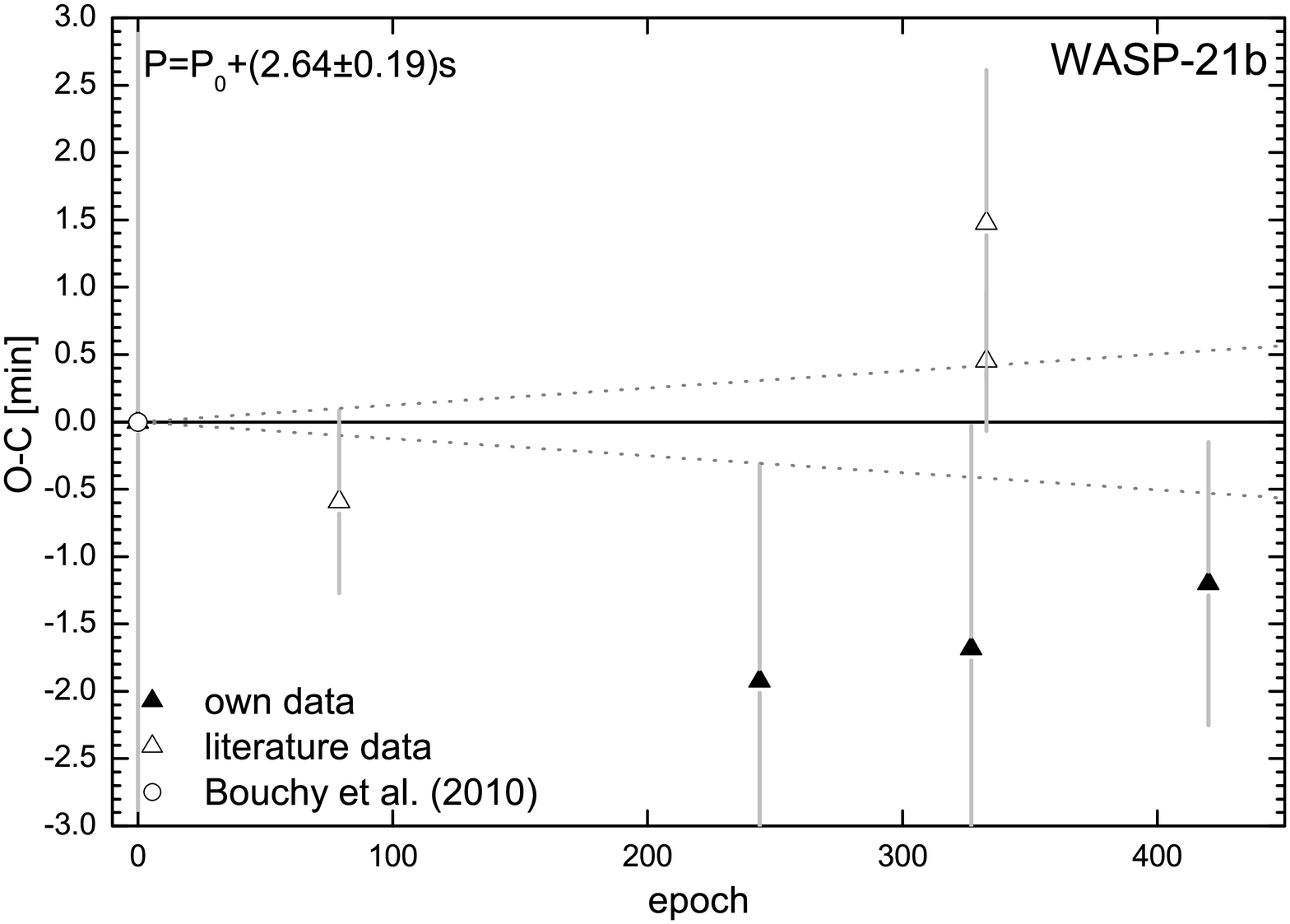}
  \includegraphics[width=0.9\columnwidth]{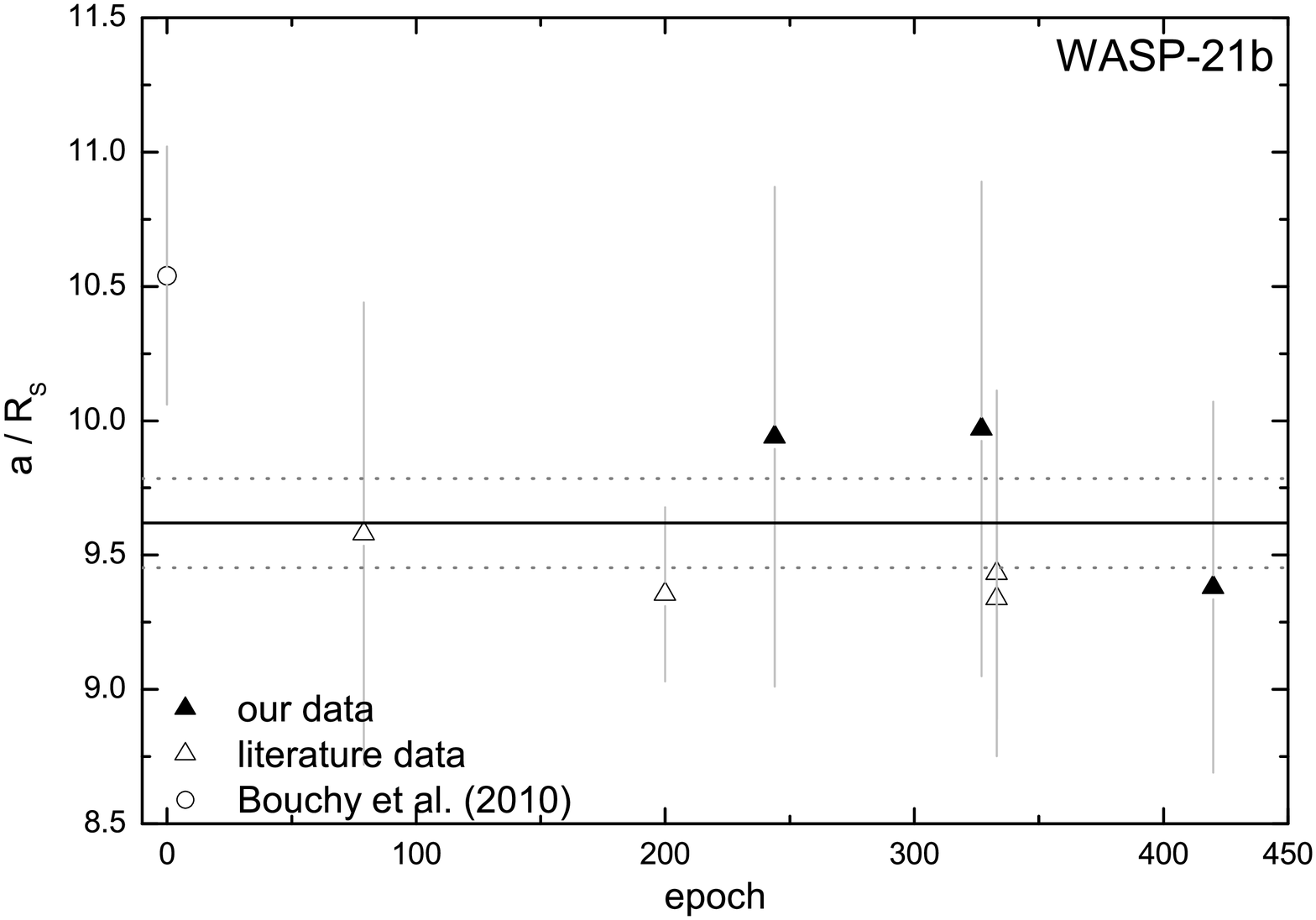}

  \includegraphics[width=0.9\columnwidth]{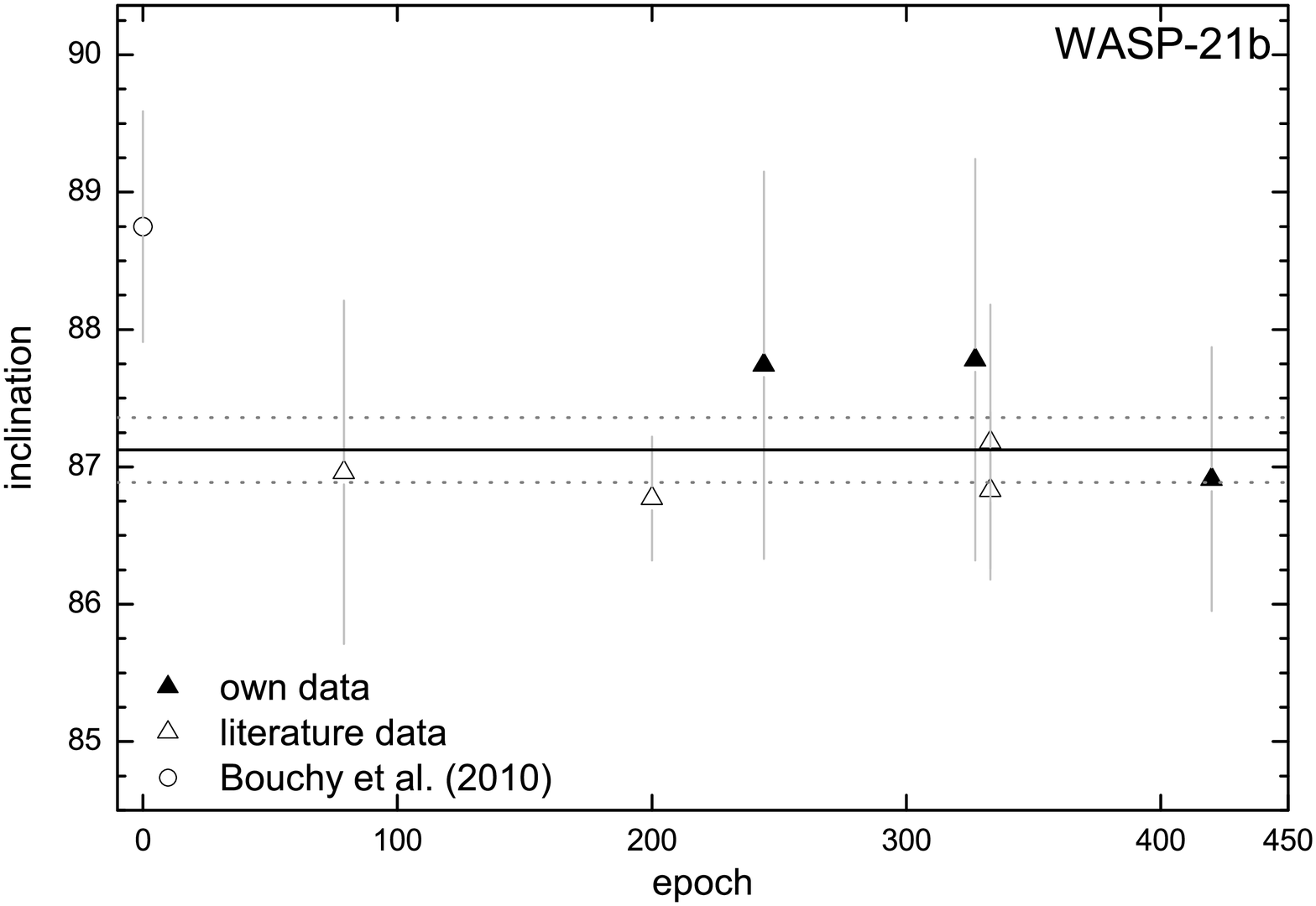}
  \includegraphics[width=0.9\columnwidth]{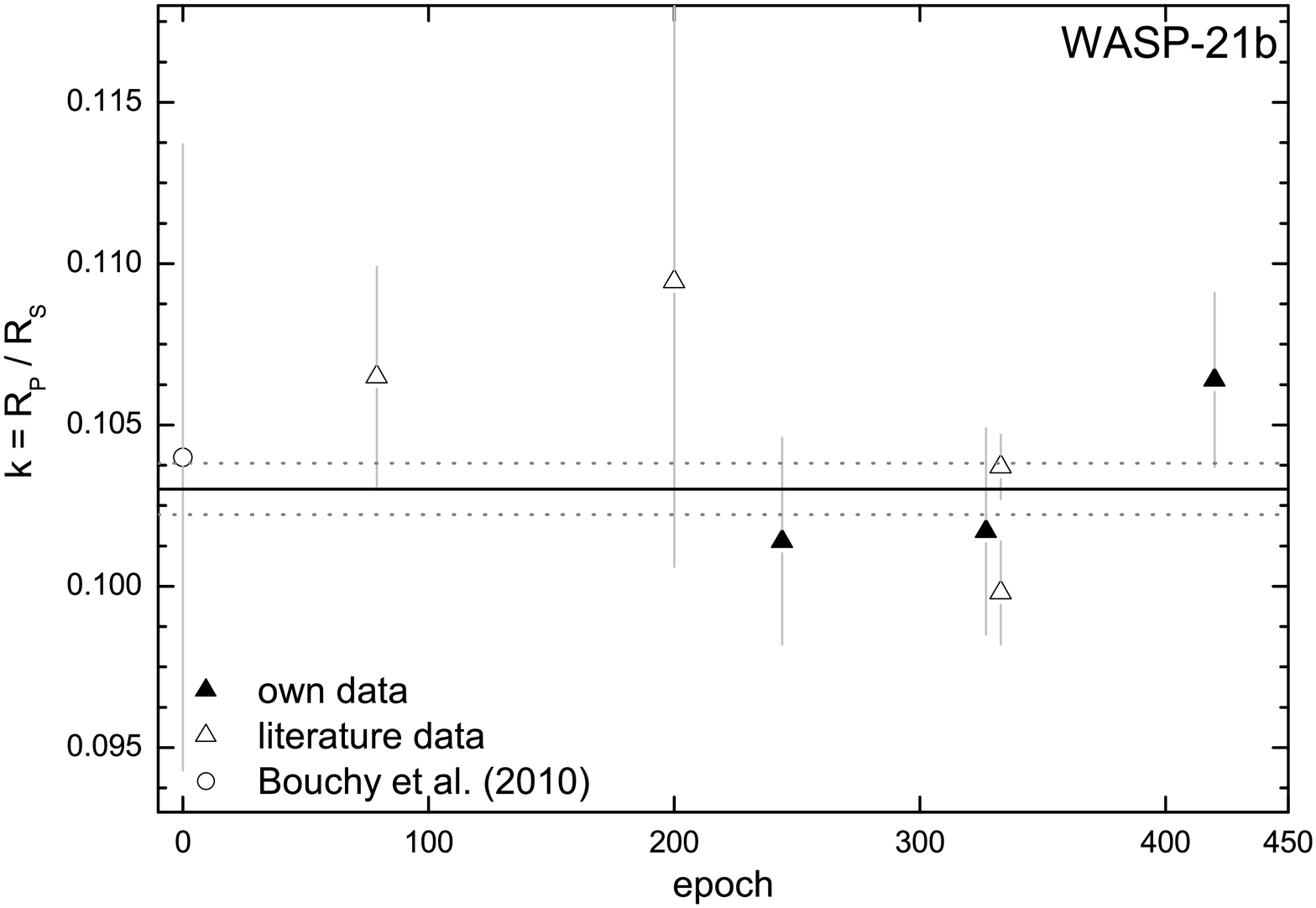}
  \caption{\textit{top:} The transit light curves obtained for WASP-21b. \textit{bottom:} 
  The present result for the WASP-21b observing campaign. All explanations are equal to 
  Fig.~\ref{fig:H18_erg}. The open circle denotes data from the discovery paper 
  of \citet{Wasp21}, open triangles denote literature data from \citet{Barros2011},
  \citet{Ciceri2013} and \citet{Southworth2012} (the latter one artificially set to epoch 200),
  filled triangles denote our data (from Swarthmore, Trebur and Calar Alto).}
  \label{fig:W21_erg}
\end{figure*}

Four transit light curves of WASP-21b are available, 
including one light 
curve from \citet{Barros2011} (see Fig.\ref{fig:W21_erg}). 
In addition, the results of the analysis of two transit
events of \citet{Ciceri2013} and one transit observation of \citet{Southworth2012}
are also taken into account. Concerning the O--C diagram,
we found that a period change of $\left(2.63\pm0.17\right)\:$s removes the linear trend
which is present in the data fitted with the initial ephemeris. 
As in the previous analyses no trend or sinusoidal variation in the system parameters can be seen.

However, regarding inclination and reverse fractional stellar radius we do see a significant 
difference between our results and the initial values published by \citet{Wasp21}.
This was also found by other authors before. As discussed in \citet{Barros2011} this result is a consequence 
of the assumption of \citet{Wasp21} that the planet host star is a main sequence star, while
\citet{Barros2011} found that the star starts evolving off the main sequence 
and thus its radius increases. 
This in turn leads to corrections of the stellar and hence planetary properties.

% \subsection{HAT-P-15b and WASP-38}
% For our targets HAT-P-15b and WASP-38b we could not get any usable light curve yet.
% 
% In case of HAT-P-15b the light curve obtained on 2011-09-28 suffers jumps in the data which are
% caused by the telescope movement in that night. This makes good fits to the data impossible.
% The observations of WASP-38b planetary transits cover only half the transit event each.
% The low number of successful transit observations is a direct consequence of the 
% long duration of the respective transits making it 
% difficult to observe a complete event.

\section{Summary}
\label{sec:Summary}

We presented the results of the transit observations of the extra solar planets
HAT-P-18b, HAT-P-19b, HAT-P-27b/WASP-40b and WASP-21b which are part of
our ongoing project on gound-based follow-up observations of 
exoplanetary transits using small to medium sized telescopes with the help of YETI network telescopes. 
During the past three years we followed these well chosen objects to refine their orbital parameters
as well as to find transit timing variations indicating yet unknown planetary companions. 
Table~\ref{tab:AllNewProperties} contains an overview of the redetermined properties, as well as the
available literature values, while Table~\ref{tab:fitResults} lists the results of the individual 
light curve fits.

In all cases we could redetermine the orbital parameters. 
Especially the period could be determined more precise than before.
So far, we can not rule out the existence of TTV signals for the planets investigated within this
study due to the limited number of available high quality data. 
Also the parameters $a/R_\textrm{s}$,
$r_\textrm{p}/r_\textrm{s}$ and inclination have been obtained and compared to the available literature data.
Despite some corrections to the literature data, we found no
significant variations within these parameters. To distinguish between
a real astrophysical source of the remaining scatter and random noise as a result of the quality of our data
more high precision transit observations would be needed.

HAT-P-18b was also part of an out-of-transit monitoring for a spread in the transit depth was 
reported in the literature that could be due to a significant variability of the transit
host star. Regarding our transit data we can not confirm the spread in transit 
depth. Looking at the quality of the literature data showing the transit depth variation, it
is very likely that this spread is of artificial nature. Thus it is not suprising that 
we did not find stellar variability larger than $\approx3.8\:$mmag. However we do see some
structures in the light curves that could be caused by spot activity on the stellar
surface.

\begin{table*}
	\caption{A comparison between the results obtained in Our analysis and the 
	literature data. 
	All epochs $T_0$ are converted to $BJD_{TDB}$.}\label{tab:AllNewProperties}
	\begin{tabular}{@{}ll@{$\,\pm\,$}ll@{$\,\pm\,$}ll@{$\,\pm\,$}ll@{$\,\pm\,$}ll@{$\,\pm\,$}l}
	\toprule
				& \multicolumn{2}{c}{$T_0\,(d)$} & \multicolumn{2}{c}{$P\,(d)$} & \multicolumn{2}{c}{$a/R_\textrm{s}$}  & \multicolumn{2}{c}{$k=R_\textrm{p}/R_\textrm{s}$} & \multicolumn{2}{c}{$i\,(^\circ)$} \\
	\midrule
	\multicolumn{11}{c}{HAT-P-18b}\\
	\midrule
	our analysis		& $2\,454\,715.022\,54$	&$0.000\,39$	& $5.508\,029\,1$	&$0.000\,004\,2$	& $17.09$		& $0.71$		& $0.136\,2$	&$0.001\,1$	& $88.79$	& $0.21$ \\
	\citet{HatP18u19}	& $2\,454\,715.022\,51$	&$0.000\,20$	& $5.508\,023$		&$0.000\,006$		& $16.04$		& $0.75$		& $0.136\,5$	&$0.001\,5$	& $88.3$	& $0.3$  \\
	\citet{Esposito2014}& $2\,455\,706.7    $	&$0.7    $		& $5.507\,978 $		&$0.000\,043 $		& $16.76$		& $0.82$		& $0.136 $		&$0.011 $	& $88.79$	& $0.25$ \\
	\midrule
	\multicolumn{11}{c}{HAT-P-19b}\\
	\midrule
	our analysis		& $2\,455\,091.535\,00$	&$0.000\,15$	& $4.008\,784\,2$	&$0.000\,000\,7$	& $12.36$		& $0.09$		& $0.137\,8$	& $0.001\,4$	& $88.51$	& $0.22$ \\
	\citet{HatP18u19}	& $2\,455\,091.534\,94$	&$0.000\,34$	& $4.008\,778$		&$0.000\,006$		& $12.24$		& $0.67$		& $0.141\,8$	& $0.002\,0$	& $88.2$	& $0.4$  \\
	\midrule
	\multicolumn{11}{c}{HAT-P-27b}\\
	\midrule
	our analysis		& $2\,455\,186.019\,91$	&$0.000\,44$	& $3.039\,580\,3$	&$0.000\,001\,5$	& $10.01$		& $0.13$		& $0.119\,2$	& $0.001\,5$	& $85.08$	& $0.07$ \\
	\citet{HatP27}		& $2\,455\,186.019\,55$	&$0.000\,54$	& $3.039\,486$		&$0.000\,012$		&\multicolumn{2}{l}{$\phantom{0}9.65\,_{-0.40}^{+0.54}$}& $0.118\,6$	&$0.003\,1$&\multicolumn{2}{l}{$84.7\phantom{0}\,_{-0.4}^{+0.7}$}\\
	\citet{Wasp40}		& $2\,455\,368.394\,76$	&$0.000\,18$	& $3.039\,572\,1$	&$0.000\,007\,8$	& $\phantom{0}9.88$		& $0.39$		& $0.125\,0$	&$0.001\,5$& \multicolumn{2}{l}{$84.98\,_{-0.14}^{+0.20}$}\\
	\citet{Sada}		& $2\,455\,186.198\,22$	&$0.000\,32$	& $3.039\,582\,4$	&$0.000\,003\,5$ 	& \multicolumn{2}{l}{$\phantom{0}9.11\,_{-1.01}^{+0.71}$} & \multicolumn{2}{l}{$0.134\,4\,_{-0.038\,9}^{+0.017\,4}$} &$84.23$&$0.88$ \\
	\citet{Brown2012}	& \multicolumn{2}{c}{--}				& $3.039\,577$		&$0.000\,006$		& \multicolumn{2}{l}{$\phantom{0}9.80\,_{-0.29}^{+0.38}$}& \multicolumn{2}{l}{$0.120\phantom{\,0}\,_{-0.007}^{+0.009}$}& $85.0$ & $0.2$\\
	\midrule
	\multicolumn{11}{c}{WASP-21b}\\
	\midrule
	our analysis		& $2\,454\,743.042\,17$	&$0.000\,65$	& $4.322\,512\,6$	&$0.000\,002\,2$	& $\phantom{0}9.62$	& $0.17$	& $0.103\,0$	& $0.000\,8$  & $87.12$	& $0.24$ \\
	\citet{Wasp21}		& $2\,454\,743.042\,6 $	&$0.002\,2 $	& \multicolumn{2}{l}{$4.322\,482\phantom{\,0}\,_{-0.000\,019}^{+0.000\,024}$}	&\multicolumn{2}{l}{$6.05\,_{-0.04}^{+0.03}$}& \multicolumn{2}{l}{$0.104\,0\,_{-0.001\,8}^{+0.001\,7}$}& \multicolumn{2}{l}{$88.75\,_{-0.84}^{+0.70}$}\\
	\citet{Barros2011}	& $2\,455\,084.520\,48$	&$0.000\,20$	& $4.322\,506\,0$	&$0.000\,003\,1$	&\multicolumn{2}{l}{$9.68\,_{-0.19}^{+0.30}$} &\multicolumn{2}{l}{$0.107\,1\,_{-0.000\,8}^{+0.000\,9}$} & $87.34$&$0.29$\\
	\citet{Ciceri2013}	& $2\,454\,743.040\,54$	&$0.000\,71$	& $4.322\,518\,6$	&$0.000\,003\,0$ 	& $9.46$ & $0.27$ & $0.1055$&$0.0023$ &$86.97$ & $0.33$ \\
	\citet{Southworth2012}&$2\,455\,084.520\,40$&$0.000\,16$	& $4.322\,506\,0$	&$0.000\,003\,1$ 	& $9.35$ & $0.34$ & $0.1095$&$0.0013$ &$86.77$ & $0.45$ \\
	\bottomrule
	\end{tabular}
\end{table*}

\begin{table*}
	\caption{The results of the induvidual fits of the observed complete transit event. 
	The $rms$ of the fit and the resultant $pnr$ are given in the last column. The table also shows the result
	for the transits with $pnr>4.5$ that are not used for redetermining the system properties.}
	\label{tab:fitResults}
	\begin{savenotes}
	{\setlength{\tabcolsep}{3.5pt}
	\begin{tabular}{lcll@{$\:\pm\:$}ll@{$\:\pm\:$}ll@{$\:\pm\:$}ll@{$\:\pm\:$}lc}
		\toprule
		date & epoch &telescope& \multicolumn{2}{c}{$T_{\textrm{mid}}-2\,450\,000\:$d}& \multicolumn{2}{c}{$a/R_\textrm{s}$} & \multicolumn{2}{c}{$k=R_\textrm{p}/R_\textrm{s}$} & \multicolumn{2}{c}{$i\,(^\circ)$} &$rms/pnr\:$(mmag)\\
		\midrule
		\multicolumn{12}{c}{HAT-P-18b}\\
		\midrule
		2011/04/21 & 174 & Trebur 1.2m     & $5\,673.419\,67$ & $0.001\,24$ & $16.4  $ & $1.4  $    & $0.139\,9$ & $0.007\,2$  & $88.52$ & $0.84$     & $3.0\,/\,3.3$\\
		2011/05/24 & 180 & Trebur 1.2m     & $5\,706.469\,93$ & $0.000\,80$ & $18.28 $ & $0.83 $    & $0.134\,3$ & $0.003\,9$  & $89.52$ & $0.58$     & $3.7\,/\,4.0$\\
		2012/05/05 & 243 & Rozhen 0.6m     & $6\,053.472\,76$ & $0.000\,84$ & $16.04 $ & $1.36 $    & $0.137\,3$ & $0.004\,7$  & $88.55$ & $0.79$     & $3.5\,/\,4.4$\\
		2012/06/07 & 249 & CA-DLR 1.2m     & $6\,086.518\,56$ & $0.001\,25$ & \multicolumn{2}{c}{--}&\multicolumn{2}{c}{--}&\multicolumn{2}{c}{--}    & $4.1\,/\,4.5$\\
		2013/04/28 & 308 & Antalya 1.0m    & $6\,411.496\,38$ & $0.000\,84$ & $15.22 $ & $1.52 $    & $0.146\,4$ & $0.006\,8$  & $87.89$ & $0.75$     & $3.9\,/\,5.0$\\
		\midrule
		\multicolumn{12}{c}{HAT-P-19b}\\
		\midrule
		2011/11/23 & 199 & Jena 0.6m       & $5\,899.283\,45$ & $0.000\,49$ & $12.56 $ & $0.34 $    & $0.136\,9$ & $0.002\,6$  & $88.20$ & $0.64$     & $2.1\,/\,2.2$\\
		2011/12/09 & 203 & Jena 0.6m       & $5\,905.318\,10$ & $0.000\,44$ & $12.29 $ & $0.35 $    & $0.136\,9$ & $0.002\,3$  & $89.05$ & $0.67$     & $2.3\,/\,2.4$\\
		2011/12/17 & 205 & CA-DLR 1.2m     & $5\,913.335\,71$ & $0.000\,34$ & $11.96 $ & $0.53 $    & $0.136\,8$ & $0.002\,7$  & $88.38$ & $0.80$     & $1.2\,/\,1.3$\\
		2014/10/04 & 460 & Jena 0.6m       & $6\,935.575\,59$ & $0.000\,55$ & $12.43 $ & $0.36 $    & $0.134\,0$ & $0.002\,6$  & $89.25$ & $0.67$     & $2.8\,/\,3.0$\\
		\midrule
		\multicolumn{12}{c}{HAT-P-27b}\\
		\midrule
		2011/04/05 & 155 & Lulin 0.4m      & $5\,657.153\,33$ & $0.001\,07$ & $10.72 $ & $1.67 $    & $0.123\,3$ & $0.008\,1$  & $85.53$ & $0.93$     & $3.4\,/\,3.8$\\
		2011/04/08 & 156 & Lulin 0.4m      & $5\,660.194\,81$ & $0.001\,16$ & $ 9.43 $ & $1.01 $    & $0.122\,8$ & $0.014\,9$  & $84.69$ & $0.81$     & $3.4\,/\,3.2$\\
		2011/05/05 & 165 & Trebur 1.2m     & $5\,687.551\,22$ & $0.000\,51$ & $ 9.83 $ & $0.56 $    & $0.115\,3$ & $0.002\,9$  & $85.07$ & $0.40$     & $1.6\,/\,1.8$\\
		2012/04/01 & 274 & Tenagra 0.8m    & $6\,018.864\,57$ & $0.002\,32$ & $ 9.65 $ & $1.63 $    & $0.119\,9$ & $0.012\,6$  & $84.13$ & $1.63$     & $5.7\,/\,5.8$\\
		2012/04/25 & 282 & Xinglong 0.6m   & $6\,043.180\,95$ & $0.001\,35$ & $ 9.89 $ & $1.67 $    & $0.118\,6$ & $0.006\,7$  & $84.83$ & $1.24$     & $4.3\,/\,5.1$\\
		2013/06/03 & 415 & Antalya 1.0m    & $6\,447.442\,68$ & $0.001\,66$ & $10.64 $ & $1.30 $    & $0.118\,4$ & $0.008\,1$  & $85.51$ & $0.94$     & $2.6\,/\,3.5$\\
		2013/06/03 & 415 & OSN 1.5m        & $6\,447.445\,71$ & $0.000\,30$ & $10.18 $ & $0.29 $    & $0.122\,4$ & $0.003\,7$  & $85.23$ & $0.21$     & $1.2\,/\,0.9$\\
		2013/06/18 & 540 & Antalya 1.0m    & $6\,827.395\,45$ & $0.002\,20$ & $10.77 $ & $1.01 $    & $0.146\,2$ & $0.014\,1$  & $85.26$ & $0.55$     & $3.1\,/\,4.0$\\
		\midrule
		\multicolumn{12}{c}{WASP-21b}\\
		\midrule
		2011/08/24 & 244 & Swarthmore 0.6m & $5\,797.734\,00$ & $0.001\,12$ & $ 9.94 $ & $0.93 $    & $0.101\,4$ & $0.003\,2$  & $87.74$ & $1.41$     & $3.3\,/\,3.1$\\
		2012/08/16 & 327 & Trebur 1.2m     & $6\,156.502\,60$ & $0.001\,15$ & $ 9.97 $ & $0.92 $    & $0.101\,7$ & $0.003\,2$  & $87.78$ & $1.46$     & $2.9\,/\,2.6$\\
		2013/09/18 & 420 & CA-DLR 1.2m     & $6\,558.496\,48$ & $0.000\,73$ & $ 9.38 $ & $0.69 $    & $0.106\,4$ & $0.002\,7$  & $86.91$ & $0.96$     & $1.6\,/\,1.4$\\
		\bottomrule
	\end{tabular}}
	\end{savenotes}
\end{table*}

\section*{Acknowledgments}

All the participating observatories appreciate the logistic and financial support of their institutions and in particular their technical workshops.
MS would like to thank all participating YETI telescopes for their observations, as well as G. Maciejewski for helpful comments on this work.
JGS, AP, and RN would like to thank the Deutsche Forschungsgemeinschaft (DFG) for support in the Collaborative Research Center Sonderforschungsbereich SFB~TR~7 ``Gravitationswellenastronomie''.
RE, MK, SR, and RN would like to thank the DFG for support in the Priority Programme SPP 1385 on the \textit{First ten Million years of the Solar System} in projects NE 515/34-1 \& -2.
RN would like to acknowledge financial support from the Thuringian government (B 515-07010) for the STK CCD camera (Jena 0.6m) used in this project.
MM and CG thank DFG in project MU 2695/13-1.
The research of DD and DK was supported partly by funds of projects DO~02-362, DO~02-85 and DDVU~02/40-2010 of the Bulgarian Scientific Foundation, as well as project RD-08-261 of Shumen University.
Wu,Z.Y. was supported by the Chinese National Natural Science Foundation grant Nos. 11373033. 
The research of RC, MH and MH is supported as a project of the Nordrhein-Westf\"alische Akademie der Wissenschaften und K\"unste in the framework of the academy programme by the Federal Republic of Germany and the state Nordrhein-Westfalen.
MF acknowledges financial support from grants AYA2011-30147-C03-01 of the Spanish Ministry of Economy and Competivity (MINECO), co-funded with EU FEDER funds, and 2011 FQM 7363 of the Consejer\'{\i}a de Econom\'{\i}a, Innovaci\'{o}n, Ciencia y Empleo (Junta de Andaluc\'{\i}a, Spain)
We also wish to thank the T\"UB\.{I}TAK National Observatory (TUG) for supporting this work through project number 12BT100-324-0 an 12CT100-388 using the T100 telescope.
MS thanks D.~Keeley, M.~M.~Hohle and H.~Gilbert for supporting the observations at the University Observatory Jena.
This research has made use of NASA's Astrophysics Data System.
This research is based on observations obtained with telescopes of the University Observatory Jena, which is operated by the Astrophysical Institute of the Friedrich-Schiller-University.
This work has been supported in part by Istanbul University under project number 39742, by a VEGA Grant 2/0143/14 of the Slovak Academy of Sciences and by the joint fund of Astronomy of
the National Science Foundation of China and the Chinese Academy of Science under Grants U1231113.

\label{lastpage}

\end{document}